\documentclass[submission,copyright,creativecommons,svgnames]{eptcs}
\providecommand{\event}{SYNT 2018} 

\usepackage[utf8]{inputenc}
\usepackage{underscore}
\usepackage{amsmath}
\usepackage{amssymb}
\usepackage{amsthm}
\usepackage{todonotes}
\usepackage{stmaryrd}
\usepackage{complexity} 
\usepackage{pgf}
\usepackage{diagbox}
\usepackage{thmtools}
\usepackage{thm-restate}
\usepackage{tikz}
\usepackage{xstring,xifthen}

\usetikzlibrary{arrows,automata,calc}

\newtheorem{problem}{Problem}
\newtheorem{definition}{Definition}
\newtheorem{lemma}{Lemma}
\newtheorem{remark}{Remark}
\newtheorem{theorem}{Theorem}

\title{Minimal Synthesis of String To String Functions From Examples}

\author{
    Jad Hamza
    \institute{LARA, EPFL, Switzerland}
    \email{jad.hamza@epfl.ch}
    \and    
    Viktor Kunčak
    \institute{LARA, EPFL, Switzerland}
    \email{viktor.kuncak@epfl.ch}
}
\def\authorrunning{J. Hamza \& V. Kunčak}

\begin{document}

\pagestyle{headings}


\newcommand{\set}[1]{\{{#1}\}}
\newcommand{\ie}{i.e.~}
\newcounter{ntodos}
\newcommand{\ctodo}[1]{
    \stepcounter{ntodos}
    \todo[inline,color=green!20]{TODO {\thentodos}: #1}
}
\newcommand\pto{\mathrel{\ooalign{\hfil$\mapstochar$\hfil\cr$\to$\cr}}}

\newcommand{\Nat}{\mathbb{N}}
\newcommand{\Var}{\mathcal{X}}

\newcommand{\ioautomaton}{f-NDMM}
\newcommand{\ioautomata}{f-NDMMs}


\newcommand{\zip}{*}
\newcommand{\emptyseq}{\varepsilon}

\newcommand{\dom}{\textit{dom}}
\newcommand{\getin}{\textit{input}}

\newcommand{\DSST}{DSST}
\newcommand{\dsst}{D}
\newcommand{\initState}{q_{\textit{init}}}
\newcommand{\update}{\rho}
\newcommand{\final}{F}
\newcommand{\semantics}[1]{\llbracket {#1} \rrbracket}
\newcommand{\val}{\mu}

\newcommand{\zero}[1]{z_{#1}}
\newcommand{\zeroconstant}{Z} 

\newcommand{\oneinthree}{One-In-Three SAT}
\newcommand{\getlang}[1]{\mathcal{L}({#1})}

\tikzstyle{tr}=[-latex,semithick]
\tikzstyle{tra}=[-latex,semithick,Red]
\tikzstyle{trb}=[-latex,semithick,Blue]
\tikzstyle{trc}=[-latex,semithick,Green]

\newcommand{\layern}[5]{
    \node[state] 
        (a#5) at (0,{#1}) { $#2$ };
    \node[state] (b#5) at (1,{#1}) { $#3$ };
    \node[state] (c#5) at (2,{#1}) { $#4$ };
} 

\newcommand{\layer}[5]{
    \node[state,accepting] 
        (a#5) at (0,{#1}) { $#2$ };
    \node[state,accepting] (b#5) at (1,{#1}) { $#3$ };
    \node[state,accepting] (c#5) at (2,{#1}) { $#4$ };
} 

\makeatletter
\providecommand\@dotsep{5}
\renewcommand{\listoftodos}[1][\@todonotes@todolistname]{%
  \@starttoc{tdo}{#1}}
\makeatother

\newcommand{\paths}[2]{
\path[tr] (a#1) edge node[left] { (a,0) } (a#2);
\path[tr] (b#1) edge node[left] { (a,0) } (b#2);
\path[tr] (c#1) edge node[left] { (a,0) } (c#2);
}

\newcommand{\prefixes}[1]{{\it Prefixes}({#1})}
\newcommand{\isPrefix}{\preceq_p}

\colorlet{c1}{Thistle}
\colorlet{c2}{blue}
\colorlet{c3}{red}

\newcommand{\pretty}[1]{
    \ifthenelse{#1 > 50}{ 
        \textcolor{Green!\the\numexpr 2*(#1 - 50)!DarkOrange}{#1}
    }{
        \textcolor{DarkOrange!\the\numexpr 2*#1!Red}{#1}
    }
}

\newcommand{\isFinal}{{\sf isFinal}}
\newcommand{\deltain}{\delta_\textit{in}}
\newcommand{\isunambiguous}{\textsf{Unambiguous}}
\newcommand{\istotal}{\textsf{Total}}
\newcommand{\corresp}{\textsf{Projection}}
\newcommand{\acceptexamples}{\textsf{AcceptExamples}}
\newcommand{\booltoint}[1]{\textsf{boolToInt}({#1})}

\newcommand{\vk}[1]{\textcolor{red}{\bf #1}}
\maketitle


\begin{abstract}

We study the problem of synthesizing string to string transformations from a set
of input/output examples. The transformations we consider are expressed using
deterministic finite automata (DFA) that read pairs of letters, one letter from
the input and one from the output. The DFA corresponding to these
transformations have additional constraints, ensuring that each input string 
is mapped to exactly one output string.

We suggest that, given a set of input/output examples, the smallest DFA
consistent with the examples is a good candidate for the transformation the user
was expecting. We therefore study the problem of, given a set of examples,
finding a minimal DFA consistent with the examples and satisfying the
functionality and totality constraints mentioned above.

We prove that, in general, this problem (the corresponding decision problem) is
$\NP$-complete. This is unlike the standard DFA minimization problem which can
be solved in polynomial time. We provide several $\NP$-hardness proofs that show
the hardness of multiple (independent) variants of the problem.

Finally, we propose an algorithm for finding the minimal DFA consistent with
input/output examples, that uses a reduction to SMT solvers. We implemented the
algorithm, and used it to evaluate the likelihood that the minimal DFA indeed
corresponds to the DFA expected by the user.

\end{abstract}



\section{Introduction}

Programming by examples is a form of program synthesis that enables users to
create programs by presenting input/output examples. In this paper, we analyze the problem of synthesizing string-to-string transformations from
examples.

We consider string transformations that can be represented by finite-state
automata, called \emph{functional non-deterministic Mealy machines}
(\ioautomaton)~\cite{mealy1955method}. \ioautomata{} output one letter for each input letter which is
read. Non-determinism refers to the fact that \ioautomata{} are allowed to have
two outgoing transitions from the same state labeled by the same input, while
functionality ensures that overall, one input string is mapped to at most one
output string. Moreover, if every input string has a corresponding output
string, the automaton is called \emph{total}.


Synthesizing an arbitrary total \ioautomaton{} consistent with input/output
examples can be solved in polynomial time, by having the \ioautomaton{} return a
default string for the inputs which are not specified in the example. The issue
with this basic approach is that the generated automaton might not be what the
user had in mind when giving the input/output examples. In other words,
input/output examples are not a complete specification, and are ambiguous.

As one of the simplest and robust criteria to rank possible solutions, we propose to synthesize a \emph{minimal} automaton consistent
with given input/output examples. For sufficiently
long input/output descriptions, the requirement
of minimality then forces the automaton to generalize
from input/output examples. This rationale is analogous
to motivation for Syntax-Guided Synthesis~\cite{alur2013}. In our case
we use automata minimality as a somewhat application-agnostic criterion. Furthermore, we can in principle leverage the insights from automata theory to improve the synthesis algorithm. Therefore, it
is interesting to understand the precise computational
complexity of such synthesis problems and to identify
directions for promising synthesis approaches. This
is the objective of our paper.

\paragraph{Complexity.}
We prove that the synthesis of minimal automata is in $\NP$, 
by showing that for a given
set of input-output examples $E$ there always exist an \ioautomaton{} consistent
with $E$ whose number of states is linear with respect to the
size of $E$. Furthermore, we show how to check in deterministic polynomial time whether a given DFA is a total \ioautomaton{} consitent with $E$.
An $\NP$ procedure can iterate for $i$ from $1$ to the aforementioned bound, 
guess a DFA of size $i$, and check that it is a total \ioautomaton{} consistent with the input/output examples.
 
We also consider the associated decision problem, which asks, given a set of
input/output examples, and a \emph{target} number of states $k$, whether there
exists a total \ioautomaton{} consistent with the examples and which has at most
$k$ states. We prove that this problem is $\NP$-hard. 

We give three distinct reductions, that apply for different variants of the
problem. First, we show that the problem is $\NP$-hard when the target number of
states is fixed to $3$ (but the input alphabet is part of the problem
description). Second, we show that the decision problem is $\NP$-hard when the
input and output alphabets are fixed (but the target number of states is part of
the problem description). 

Third, we study a variant of the problem for \emph{layered} automata that
recognize only words of some fixed length. The name \emph{layered} comes from
the fact that their states can be organized into layers that recognize only
words of a certain length. We prove that the problem is still $\NP$-hard in 
that setting, despite the fact that these automata have no cycles.

\paragraph{Algorithm.}
We provide a reduction to the satisfiability of a logical formula. We implement
our reduction, and link it to the Z3 SMT solver. We evaluate our tool and show
it can successful recover simple relations on strings from not too many examples
(but scales to many examples as well). We also 
evaluate the ability of our algorithm to recover a random automaton from
a sample set of input-output examples. Our experiments suggest that it is better to give a large number of small examples, rather than a small number of large
examples. Moreover, to improve the chance that our algorithm finds a particular
automaton, the examples given should generally be at least as long as the number
of states.

\paragraph{Contributions} of this paper are the following:
\begin{itemize}
\item $\NP$-hardness proofs for the decision problem 
    (Sections~\ref{sec:fixedstates} and~\ref{sec:others}),
\item Proof that the minimization problem can be solved in $\NP$
    (Section~\ref{section:easiness}),
\item A reduction from the minimization problem to a logical formula that can be
    handled by SMT solvers (Section~\ref{section:algorithm}),
\item An implementation of this reduction
and experiments that evaluate the likelihood that minimization finds the
    automaton the user has in mind (Section~\ref{section:experiments}).
\end{itemize}

Due to space constraints, some proofs are deferred to the Appendix.


\section{Notation}

An \emph{alphabet} $\Sigma$ is a non-empty finite set.
Given a natural number $n \in \Nat$, we denote by 
$\Sigma^n$ the set of sequences (or words) of $n$ symbols of $\Sigma$. 
We denote by $\Sigma^*$ the set of finite sequences 
$\bigcup_{n \geq 0} \Sigma^n$.
For $u \in \Sigma^*$, $|u|$ denotes the length of 
the sequence $u$.
A set of words is called a \emph{language}.

A \emph{non-deterministic finite automaton (NFA)} $A$ is a tuple
$(\Sigma,Q,\initState,\delta,F)$ where $\Sigma$ is an alphabet, $Q$ is the
finite set of states, $\initState \in Q$ is the initial state, $\delta \subseteq
Q \times \Sigma \times Q$ is the transition function, and $F \subseteq Q$ is the
set of \emph{accepting} states. We denote by $\getlang{A}$ the language accepted
by $A$, i.e. the set of words for which there exists an \emph{accepting run} in
$A$. By an abuse of notation, the set $\getlang{A}$ is sometimes denoted by $A$.

An NFA $A$ is \emph{unambiguous} (denoted \emph{UFA}) 
if every word in $\Sigma^*$ has at most one accepting run in $A$.
An NFA is \emph{deterministic} (denoted \emph{DFA}) if for every $q_1 \in Q$,
$a \in \Sigma$, there exists a unique $q_2 \in Q$ such that 
$(q_1,a,q_2) \in \delta$.
The \emph{size} of an NFA $A$ is its number of states, and is denoted 
$|A|$.

Let $\Sigma$ and $\Gamma$ be two alphabets.
For $u \in \Sigma^{n}$ and $v \in \Gamma^n$ where 
$u = u_1 \dots u_n$, $v = v_1 \dots v_n$, we denote by $u \zip v$ the 
sequence in $(\Sigma \times \Gamma)^n$ 
where $u \zip v = (u_1, v_1) \dots (u_n, v_n)$.
Note that the operator $\zip$ is well defined only when $|u| = |v|$.



Given two words $u, v \in \Sigma^*$, we denote by $u \isPrefix v$ the fact
that $u$ is a prefix of $v$. Moreover, $\prefixes{v}$ denotes the set of 
prefixes of $v$, that is $\prefixes{v} = \set{u \ |\ u \isPrefix v}$.


\section{Functional Non-Deterministic Mealy Machines}

We consider two alphabets, an \emph{input alphabet} $\Sigma$ and an \emph{output
alphabet} $\Gamma$. A \emph{functional non-deterministic Mealy machine
(\ioautomaton)} is a DFA $A$ over $\Sigma \times \Gamma$ satisfying: for all $u
\in \Sigma^*$, $v_1,v_2 \in \Gamma^*$, if $u \zip v_1 \in \getlang{A}$ and $u
\zip v_2 \in \getlang{A}$, then $v_1 = v_2$. 

Note here that we model \ioautomata{} with \emph{deterministic} finite automata.
The determinism refers to the fact given a state, an input letter and an output
letter, there is at most one outgoing transition labeled by those letters. On
the other hand, the non-determinism in the \ioautomaton{} refers to the fact
that given one state and one input letter, there might be multiple outgoing
transitions, each one labeled with a distinct output letter.

Due to the functionality restriction described above, an \ioautomaton{} $A$
defines a partial function $\bar A \subseteq \Sigma^* \times \Gamma^*$, which is
defined for $u \in \Sigma^*$ only when there exists (a unique) $v \in \Gamma^*$
such that $u \zip v \in \getlang{A}$. This unique word $v$ is denoted by $A(u)$.
An \ioautomaton{} $A$ is called \emph{total} if the partial function $\bar A$ is
total.
For a set $E \subseteq \Sigma^* \times \Gamma^*$
we say that an \ioautomaton{} $A$ is consistent with $E$ if 
$E \subseteq \bar A$.

\begin{problem}
\label{pb:sample}
Let $E \subseteq (\Sigma \times \Gamma)^*$ be a set of input/output examples.

Find a total \ioautomaton{}, consistent with $E$ (if it exists), whose size
 is
minimal (among all total \ioautomata{} consistent with $E$).
\end{problem}

We also investigate the following corresponding decision problem.

\begin{restatable}{problem}{decsample}
\label{pb:decsample}
Let $E \subseteq (\Sigma \times \Gamma)^*$ 
be a set of input/output examples, and let $n \in \Nat$.

Does there exist a total \ioautomaton{}, consistent with $E$, with size at most
$n$?
\end{restatable}

When stating complexity results, we consider that the size of the problem is 
the sum of the sizes of each word in $E$, plus the size of $n$. Our hardness 
results holds even when $n$ is represented in unary, while our proofs that 
Problems~\ref{pb:sample} and \ref{pb:decsample} belong to $\NP$ hold even 
when $n$ is represented in binary.


\subsection{Summary of the Complexity Results}
\label{sec:summary}

Table~\ref{table:summary} summarizes the various complexity results proved in
this paper. As far as we know, the problem is open when the input alphabet has
size one, i.e.~$|\Sigma| = 1$. On the other hand, when $|\Gamma| = 1$, the
problem becomes trivial as the minimal total \ioautomaton{} consistent with
given input/output examples always has a single state with a self-loop.

Layered \ioautomata{} are defined in 
Section~\ref{sec:layered}, and are \ioautomata{} that only recognize words of 
some particular length. Even in that setting, the problem is $\NP$-complete.


\begin{table}
  
  \begin{tabular}{l|l|l}
  Problem         &    Layered \ioautomata    &     \ioautomata \\\hline
  Problem~\ref{pb:decsample}  & $\NP$-complete & $\NP$-complete \\
  With $|\Gamma| = 2$, $n = 3$, $|E| = 1$
      & $O(1)$  (Remark~\ref{remark:layered})
      & $\NP$-complete (Sect.~\ref{sec:fixedstates}) \\
  With $|\Sigma| = 3$, $|\Gamma| = 2$ 
     & $\NP$-complete (Sect.~\ref{sec:layered}) 
     & $\NP$-complete (Sect.~\ref{sec:fixedalphabets})\\
  With $|\Sigma| = 3$, $|\Gamma| = 2$, $|E| = 1$ 
     & $O(1)$ (Remark~\ref{remark:layered})
     & $\NP$-complete (Sect.~\ref{sec:fixedalphabets})\\
  When $\Sigma$, $\Gamma$ and $n$ are fixed & 
    in $P$ (Remark~\ref{remark:fixed}) & 
    in $P$ (Remark~\ref{remark:fixed}) \\ 
  \end{tabular}

    \caption{Summary of the complexity results
  \label{table:summary}}
  
\end{table}


\section{Preliminaries for the $\NP$-hardness proofs}

In Sections~\ref{sec:fixedstates}, \ref{sec:fixedalphabets},  and
\ref{sec:layered}, we prove $\NP$ hardness results for
Problem~\ref{pb:decsample} and variants. These hardness results carry directly
over to Problem~\ref{pb:sample}. Indeed, any algorithm for solving
Problem~\ref{pb:sample} can be adapted to solve Problem~\ref{pb:decsample}.

Our proofs rely on reductions from a variant of the boolean satisfiability
problem (SAT), called One-In-Three SAT. In all reductions, our goal is to build
from an instance $\varphi$ of One-In-Three SAT a set of input/output examples
such that $\varphi$ is satisfiable if and only if there exists a total
\ioautomaton{} consistent with the examples (and satisfying the constraints of
the minimization problem at hand). 


\begin{figure}[ht]

  \centering

  \begin{tikzpicture}[xscale=3,yscale=2.2]
  \node[state,accepting] (a) at (0,0) {$q_0$};
  \node[state,accepting] (b) at (1,0) {$q_1$};
  \node[state,accepting] (c) at (3,0) {$q_{n-1}$};

  \path[tr] (-0.3,0) edge (a);
  \path[tr] (a) edge node[above] {(a,0)} (b);
  \path[tr,dashed] (b) edge node[above] {(a,0)} (1.7,0);
  \path[tr,dashed] (2.3,0) edge node[above] {(a,0)} (c);
  \path[tr] (c) edge[bend right=35] node[above] {(a,1)} (a);
  \end{tikzpicture}

  \caption{The form of automata that have an $(a,0,1)$-loop.}
  \label{fig:a01loop}

\end{figure}
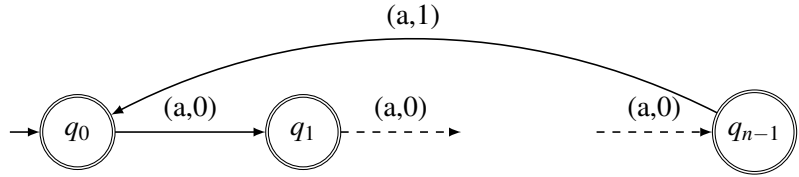

Our strategy for these reductions is to give input/output examples that
constrain the shape of \emph{any} total \ioautomaton{} consistent with these
examples. We give input/output examples that ensure that any total 
\ioautomaton{} consistent with the examples must have certain transitions, and
cannot have certain other transitions.

For example, in Sections~\ref{sec:fixedstates} and \ref{sec:fixedalphabets}, we
provide input/output examples that restrict the shape of any solution to be of
the form given in Figure~\ref{fig:a01loop}. Then, knowing that any solution must
have this shape, we give additional examples that correspond to our encoding of
$\varphi$.

We first give a formal definition for automata that are of the shape of
the automaton given in Figure~\ref{fig:a01loop}.

\begin{definition}
Let $A = (\Sigma \times \Gamma, Q, \initState, \delta, F)$ 
be an \ioautomaton{} with $n \in \Nat$ states, $n \geq 1$.
We say that $A$ \emph{has an $(a,0,1)$-loop} if 
$a \in \Sigma$, and $0,1 \in \Gamma$, $0 \neq 1$, and 
the states $Q$ of $A$ can be ordered in a sequence 
$q_0,\dots,q_{n-1}$ such that:
\begin{itemize}
\item $\initState = q_0$,
\item for every $0 \leq i < n-1$, $(q_i,(a,0),q_{i+1}) \in \delta$, 
\item $(q_{n-1},(a,1),q_0) \in \delta$,
\item $F = Q$,
\item 
    there are no transitions in $\delta$ labeled with letter $a$ other 
    than the ones mentioned above.
\end{itemize}
\end{definition}

This lemma, used in Theorems~\ref{th:fixedstates-oneexample} and
\ref{th:fixedalphabet-oneexample}, shows that we can give an input/output
example that forces automata to have an $(a,0,1)$-loop. The idea is to give a
long example that can only be recognized if the total \ioautomaton{} has an
$(a,0,1)$-loop.

\begin{lemma}
\label{lemma:a01loop} Let $A = (\Sigma \times \Gamma, Q, \initState, \delta, F)$
be a total \ioautomaton{} with $n$ states, $n \geq 1$. Let $u$ and $v$ be two
words such that:
\[
    A(a^{2n} \cdot u) = 0^{n-1}10^{n-1}1 \cdot v.
\]

Then $A$ has an $(a,0,1)$-loop.
\end{lemma}

\begin{proof}
Consider the run of $a^{2n} \zip 0^{n-1}10^{n-1}1$ in $A$, of the 
form: 
\[
    \initState = q_0 
        \xrightarrow{(a,0)} q_1 
        \xrightarrow{(a,0)} \dots 
        \xrightarrow{(a,0)} q_{n-1}
        \xrightarrow{(a,1)} q_n
        \xrightarrow{(a,0)} q_{n+1}
        \dots
        \xrightarrow{(a,0)} q_{2n-1}
        \xrightarrow{(a,1)} q_{2n}
\]
where for all $0 \leq i \leq 2n$, $q_i \in Q$. By assumption, we know
that from state $q_{2n}$, $A$ accepts $u \zip v$.

We want to prove that:
\begin{enumerate}
\item \label{enum:distinct}
  the states $q_0$ to $q_{n-1}$ are all distinct, and
\item \label{enum:wrap}
  $q_n = q_0$, and
\item \label{enum:clean}
  there are no transitions labeled by $a$ except the ones from the run 
  above, and
\item \label{enum:allfinal}
  $F = Q$.
\end{enumerate}

Note that this entails that $q_i = q_{n+i}$ for all $0 \leq i \leq n$.

$(\ref{enum:distinct})$ 
Assume by contradiction that there exists $0 \leq i < j \leq n-1$ such that 
$q_i = q_j$.
Since $A$ only has $n$ states, we know that there exists 
$n \leq k < l \leq 2n$ such that $q_k = q_l$.
We consider two cases, either $l < 2n$, or $l = 2n$.
If $l < 2n$, then the following words are accepted by $A$, leading 
to a contradiction to the output-uniqueness property of \ioautomata.
\begin{itemize}
\item 
  $a^{2n-j+i-l+k+(j-i)(l-k)} \cdot u \zip 
  0^{n-1-j+i} 1 0^{n-1-l+k+(j-i)(l-k)} 1 \cdot v$, by going through\\
  $q_0 \dots q_i q_{j+1} \dots q_{k-1} (q_k \dots q_{l-1})^{j-i} q_l \dots q_{2n} \dots$,
\item 
  $a^{2n-j+i-l+k+(j-i)(l-k)} \cdot u \zip 
  0^{n-1-j+i+(j-i)(l-k)} 1 0^{n-1-l+k} 1 \cdot v$, by going through\\
  $q_0 \dots q_{i-1} (q_i \dots q_{j-1})^{l-k} q_{j} \dots q_k q_{l+1} \dots q_{2n} \dots$.
\end{itemize}

Similarly, if $l = 2n$, the following words are accepted by $A$, again 
leading to a contradiction.
\begin{itemize}
\item 
  $a^{2n-j+i-l+k+(j-i)(l-k)} \cdot u \zip 
  0^{n-1-j+i} 1 0^{n-l+k} (0^{l-k-1}1)^{(j-i)} \cdot v$,
\item 
  $a^{2n-j+i-l+k+(j-i)(l-k)} \cdot u \zip 
  0^{n-1-j+i+(j-i)(l-k)} 1 0^{n-l+k} \cdot v$.
\end{itemize}

We conclude that the states $q_0$ to $q_{n-1}$ are all distinct.

$(\ref{enum:wrap})$ 
Since the states $q_0$ to $q_{n-1}$ are all distinct, we know that $q_n = q_i$
for some $0 \leq i \leq n-1$. 
Assume by contradiction that $0 < i$.
By doing the same case analysis as above (either $l < 2n$, or $l = 2n$), 
we again find contradictions to the output-uniqueness property of $A$.

$(\ref{enum:clean})$ 
Assume by contradiction that there exists $i \neq j$ with 
$0 \leq i,j \leq n-1$ and 
$b \in \Gamma$ such that $\delta(q_i, (a,b)) = q_j$ and this transition 
is different than the transitions from the run above.

If $i < j$, then there is an alternative loop
$q_i,q_j,q_{j+1},\dots,q_{n-1},q_0,q_1,\dots,q_i$ containing 
$n-j+i+1$ transitions labeled by $a$. In particular, this means that 
the word $a^{n + n(n-j+i+1)}$ has two different outputs in $A$.
The first one is obtained by going from $q_0$ to $q_i$, 
taking the alternative loop $n$ times, and then going from $q_i$ to 
$q_0$ using the $(a,0,1)$-loop.
The second is obtained by going from $q_0$ to $q_i$,
taking the $(a,0,1)$-loop $(n-j+i+1)$ times, and then going from 
$q_i$ to $q_0$ using the $(a,0,1)$-loop.
This contradicts the output-uniqueness property of $A$.

A similar reasoning applies when $j < i$, by using 
$q_i, q_j, q_{j+1}, \dots, q_i$ as the alternative loop.

$(\ref{enum:allfinal})$ Due to the previous property, the only run labeled whose
input is $a^i$ for $0 \leq i \leq n-1$ is the one going through
$q_0,q_1,\dots,q_i$ in the $(a,0,1)$-loop. This entails that for $0 \leq i \leq
n-1$, $q_i$ is final and $F = Q$.
\end{proof}

The following lemma states that multiple input/output examples may be encoded
into just one example for \ioautomata{} that have an $(a,0,1)$-loop.

\begin{lemma}
\label{lemma:splitexamples}
Let $A = (\Sigma \times \Gamma, Q, \initState, \delta, F)$ 
be an \ioautomaton{} with an $(a,0,1)$-loop. 
Let $u,v \in \Sigma^*$ and $u',v' \in \Gamma^*$ such that:

\[
    A(u \cdot a \cdot v) = u' \cdot 1 \cdot v'.
\]
Then $A(u \cdot a) = u' \cdot 1$ and $A(v) = v'$.
\end{lemma}

\begin{proof}
Using Lemma~\ref{lemma:a01loop}, we know that $A$ has an $(a,0,1)$-loop.
Therefore, the only transition labeled by $(a,1)$ is the one leading to the
initial state. Therefore, after reading $(u \cdot a) \zip (u' \cdot 1)$, $A$
must be in the initial state. This entails that $A(u \cdot a) = u' \cdot 1$ and
$A(v) = v'$.
\end{proof}


\section{NP-Hardness of the Minimization Problem with one Input/Output Example
and Fixed Number of States}
\label{sec:fixedstates}


\begin{figure}
\centering

\begin{tikzpicture}[xscale=5,yscale=3]

\node[state,accepting] (a) at (0,0) {$q_0$};
\node[state,accepting] (b) at (1,0) {$q_1$};
\node[state,accepting] (c) at (2,0) {$q_2$};

\path[tr] (-0.3,0) edge (a);
\path[tr] (a) edge[bend left=25] node[above] {(a,0)} (b);
\path[tr] (b) edge[bend left=25] node[above] {(b,0)} (a);
\path[tr] (b) edge[bend left=0] node[above] {(a,0)} (c);
\path[tr] (c) edge[bend left=25] node[below] {(b,1)} (a);
\path[tr] (a) edge[loop above] node[above] {(b,0)} (a);
\path[tr] (c) 
    edge[bend right=55] 
    node[above] {(a,1)} 
    (a);

\end{tikzpicture}

\caption{
    \ioautomaton{} used in the proof of Theorem~\ref{th:fixedstates-oneexample}.
}
\label{fig:base}
\end{figure}
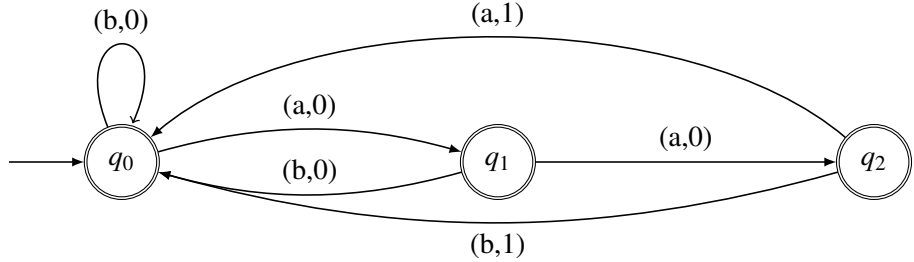 

We prove the $\NP$-hardness of Problem~\ref{pb:decsample} by reducing the
\oneinthree{} problem to it. This $\NP$-hardness proof holds even when the
target number of states for minimization is fixed to $3$, the size of the output
alphabet is fixed to $2$, and there is single input/output example.

\begin{problem}[\oneinthree]
Given a set of variables $V$ and a set of clauses $C \subseteq V^3$, 
does there exist an assignment $f: V \rightarrow \set{\bot,\top}$ such that
for each $(x,y,z) \in C$, exactly one variable out of $x$, $y$, $z$,
evaluates to $\top$ through $f$.
\end{problem}

\begin{restatable}{theorem}{fixedstates}
\label{th:fixedstates-oneexample}
Problem~\ref{pb:decsample} is $\NP$-hard when the number of states is fixed,
the output alphabet is fixed, and there is a single input/output example.
\end{restatable}

\begin{proof}
  Consider an instance $\varphi$ of \oneinthree{}, with a set of variables $V$,
  and a set of clauses $C \subseteq V^3$. We reduce \oneinthree{} to
  Problem~\ref{pb:decsample} as follows. We define $\Sigma = V \cup \set{a,b}$,
  where $a$ and $b$ are fresh symbols and $\Gamma = \set{0,1}$. Moreover, we define $n =
  3$ (fixed number of states).
  
  Then, we define $E = \set{w}$ where $w$ is one input/output example made of
  the concatenation of all the following words (the word $aaaaaa \zip 001001$
  must go first in the concatenation, but the other words can be concatenated in
  any order):
  \begin{itemize}
  \item $aaaaaa \zip 001001$,
  \item $baaa \zip 0001$,
  \item $abaaa \zip 00001$,
  \item $aabaaa \zip 001001$,
  \item $xbaaa \zip 00001$ for all $x \in V$,
  \item $xxxaaa \zip 000001$ for all $x \in V$,
  \item $axxxaa \zip 000001$ for all $x \in V$,
  \item $aaxxxa \zip 000001$ for all $x \in V$,
  \item $xyzaa \zip 00001$ for all $(x,y,z) \in C$.
  \end{itemize}
  We prove that $\varphi$ has a satisfying assignment if and only if 
  there exists a total \ioautomaton{} $A$, consistent with $E$, and with 
  (at most) $3$ states.
  
  $(\Rightarrow)$
  Let $f: V \rightarrow \set{\bot,\top}$ be a satisfying assignment for 
  $\varphi$.
  We define $A = (\Sigma\times\Gamma,Q,\initState,\delta,F)$ following 
  Figure~\ref{fig:base} with $Q = F = \set{q_0,q_1,q_2}$ and 
  $\initState = q_0$.
  The transitions involving $a \in \Sigma$ in $A$ are:
      $(q_0,(a,0),q_1),(q_1,(a,0),q_2) \in \delta$, and 
      $(q_2,(a,1),q_0) \in \delta$.

  Then, for each $x \in V$, if $f(x) = \top$, we add three 
  transitions in $\delta$, called \emph{forward transitions}:
  $(q_0, (x,0), q_1)$, 
  $(q_1, (x,0), q_2)$, and
  $(q_2, (x,0), q_0)$. 
  If $f(x) = \bot$, we add three transitions as well, called 
  \emph{looping transitions}:
  $(q_0, (x,0), q_0)$, 
  $(q_1, (x,0), q_1)$, and
  $(q_2, (x,0), q_2)$. 
  
  $A$ is a total \ioautomaton{}, since all states are final, and for every 
  state and every input in $\Sigma$, there is a unique outgoing transition
  labeled by this input (and some output in $\Gamma$). Moreover, we can verify
  that $A$ is consistent with the input/output example $w$.
  
  $(\Leftarrow)$
  Let $A = (\Sigma\times\Gamma,Q,\initState,\delta,F)$ be a total \ioautomaton{}
  with $3$ states, and consistent with $E$. Our proofs goes as follows.
  First, using Lemma~\ref{lemma:a01loop} and Lemma~\ref{lemma:splitexamples},
  we deduce that $A$ must have an $(a,0,1)$-loop, and must accept all the 
  individual words that constitute the concatenation $w$.
  Then, using the facts that
  $A(baaa) = 0001$,
  $A(abaaa) = 00001$,
  $A(aabaaa) = 001001$, we deduce that $A$ must contain the transitions 
  present in Figure~\ref{fig:base}, and no other transitions labeled by 
  $b$.

  Then, for each variable $x \in V$, using the facts that $A(xbaaa) = 00001$ and
  $A(xxxaaa) = A(axxxaa) = A(aaxxxa) = 000001$,
  we show that $x$ must either have looping transitions, 
  or forward transitions, as described in the first part of the proof.
  We then use this fact to define $f$ that assigns 
  $\top$ to variables that have forward transitions, and $\bot$ to variables 
  that have looping transitions.
  
  Finally, for each clause $(x,y,z) \in C$, and using 
  $A(xyzaa) = 00001$, we deduce that exactly one variable out of 
  $x$, $y$ and $z$ must have forward transitions, and conclude that 
  $f$ is a satisfying assignment for $\varphi$.
  
  We now give more details for each step of the proof.
  Our first goal is to prove that $A$ must contain the transitions
  given in Figure~\ref{fig:base}.
  Since $A(baaa) = 0001$, we know that after reading $(b,0)$,
  $A$ must be in state $q_0$, and therefore there exists 
  a transition $(q,0,(b,0),q_0) \in \delta$.
  Using $A(abaaa) = 00001$ and $A(aabaaa) = 001001$ respectively, we deduce that
  there exist transitions $(q,1,(b,0),q_0)$ and $(q,2,(b,1),q_0)$ in $\delta$.
  Using the output-uniqueness property of $A$, we can verify that there can be no
  other transitions labeled by $b$ in $A$.

  Our next goal is to prove that for each variable $x \in V$, 
  $x$ must either have looping transitions or forward transitions.
  
  Since $xbaaa \zip 00001 \in A$ and the only transitions labeled by $(b,0)$ are
  the ones from states $q_0$ and $q_1$, we deduce that from the initial state, 
  reading $(x,0)$ must lead either to $q_0$ or $q_1$, and therefore 
  there should either exist
  a transition $(q_0,(x,0),q_1) \in \delta$ or a 
  transition $(q_0,(x,0),q_0) \in \delta$.
  
  Assume $(q_0,(x,0),q_1) \in \delta$. In that case, we prove that $x$ has 
  forward transitions, in the sense that 
  there are transitions $(q_1,(x,0),q_2)$ and $(q_2,(x,0),q_0)$ in $\delta$.
  We know $xxxaaa \zip 0000001 \in A$. 
  Since the only state from which the word $aaa \zip 001$ is accepted
  is $q_0$, the automaton $A$ must end in $q_0$ after reading 
  $xxx \zip 000$.
  Moreover, since $(q_0,(x,0),q_1) \in \delta$, we know $A$ 
  ends in state $q_1$ after reading $(x,0)$ in the initial state.
  Therefore, when reading $xx \zip 00$ from state $q_1$, $A$ must end in state 
  $q_0$. The only way this is possible is by having transitions
  $(q_1,(x,0),q_2)$ and $(q_2,(x,0),q_0)$ in $\delta$.
  
  The other case we consider is when $(q_0,(x,0),q_0) \in \delta$. 
  Here, we want to prove that $x$ has looping transitions, with
  $(q_1,(x,0),q_1)$ and 
  $(q_2,(x,0),q_2)$ in $\delta$.
  We know $axxxaa \zip 000001 \in A$.
  The only state from which $aa \zip 01$ can be accepted is $q_1$.
  Moreover, $A$ ends in state $q_1$ after reading $(a,0)$.
  Therefore, $A$ must go from state $q_1$ to $q_1$ by reading 
  $xxx \zip 000$. Due to the self-loop $(q_0,(x,0),q_0) \in \delta$,
  the only possibility for this is to have a loop
  $(q_1,(x,0),q_1) \in \delta$. Similarly, using $aaxxxa \zip 000001 \in A$, 
  we deduce there is a loop $(q_1,(x,0),q_1) \in \delta$.

  Overall, we have shown that each variable $x \in V$ either has forward
  transitions, or looping transitions. We now define the assignment $f$ that 
  assigns $\top$ to variables that have forward transitions, and $\bot$ to 
  variables that have looping transitions.
  Let $(x,y,z) \in C$. We know $xyzaa \zip 00001 \in A$.
  The only state from which $aa \zip 01$ can be accepted is $q_1$. 
  Therefore, $A$ must end in state $q_1$ after reading $xyz \zip 000$.
  The only way for this to be the case is that exactly one of 
  $x$, $y$, $z$ has forward transitions, while the two others have looping 
  transitions.
\end{proof}
  

\section{NP-Hardness Proofs for Other Variants}
\label{sec:others}

In this section, we give two other $\NP$-hardness proofs, that cover instances
of the problem which are not comparable to the ones treated in 
Section~\ref{sec:fixedstates}.

These proofs also follow the idea of reducing from the One-in-Three SAT problem,
but require new encodings. For space constraints, the proofs are deferred to the
appendix.

\subsection{NP-Hardness of the Minimization Problem with One Input/Output Example 
and Fixed Alphabets}
\label{sec:fixedalphabets}

Our second $\NP$-hardness proof holds for the case where the sizes of both input
and output alphabets are fixed, and there is a single input/output example. When
the input and output alphabets are fixed, we can no longer use the encoding
given in the previous section, where we could associate to each variable of the
SAT formula a letter in our alphabet. Instead, we here rely on the fact that the
target number of states is not fixed. As such, this theorem is complementary 
to Theorem~\ref{th:fixedstates-oneexample} (see Appendix~\ref{app:fixedalphabets} for the proof).

\begin{restatable}{theorem}{fixedalphabets}
\label{th:fixedalphabet-oneexample}
Problem~\ref{pb:decsample} is $\NP$-hard when the alphabets $\Sigma$ and 
$\Gamma$ are fixed, and there is a single input/output example.
\end{restatable} 

\begin{remark}
  \label{remark:fixed} Note that if the input and output alphabets as well as
  the target number of states are fixed, then Problem~\ref{pb:decsample} can be
  solved in polynomial time. The reason is that when all these parameters are
  constants, then there is only a constant number of \ioautomata{} to explore.
\end{remark}


\subsection{NP-Hardness of the Minimization Problem for Layered Automata}
\label{sec:layered}

In this section, we cover automata that only recognize words of the same length.
An NFA $A = (\Sigma,Q,\initState,\delta,F)$ is said to be \emph{$l$-layered}
for $l \in \Nat$ if $A$ only accepts words of length $l$, \ie 
$\getlang{A} \subseteq \Sigma^l$.
An $l$-layered \ioautomaton{} 
$A = (\Sigma \times \Gamma,Q,\initState,\delta,F)$ 
is called \emph{$l$-total} if the domain of the function associated with $A$ 
is $\Sigma^l$.

We then adapt Problem~\ref{pb:decsample} for this setting.

\begin{problem}
\label{pb:layered}
Let $\Sigma$ be an input alphabet, $\Gamma$ an output alphabet, and 
$l \in \Nat$.
Let $u_1 \zip v_1$, ..., $u_k \zip v_k$ be a set of input/output examples,
with $u_i \in \Sigma^l$ and $v_i \in \Gamma^l$ for all $1 \leq i \leq k$.
Let $n \in \Nat$.

Does there exist an $l$-layered and $l$-total \ioautomata{} that accepts
$u_i \zip v_i$ for all $1 \leq i \leq k$, and that has at most $n$ states.
\end{problem}

The following theorem (proof in Appendix~\ref{app:layered}) proves that
Problem~\ref{pb:layered} is $\NP$-hard, even when the alphabets are fixed. In
this theorem, we can no longer rely on Lemmas~\ref{lemma:a01loop} and
\ref{lemma:splitexamples}, since layered automata cannot contain cycles.
Instead, we have to use multiple input/output examples in our encoding. 
 
\begin{restatable}{theorem}{layered}
\label{th:layered}
Problem~\ref{pb:layered} is $\NP$-hard when the alphabets $\Sigma$ and $\Gamma$ 
are fixed.
\end{restatable}

\begin{remark}
  \label{remark:layered} When there is a single input/output example,
  Problem~\ref{pb:layered} can be solved in polynomial time. The reason is that,
  in a layered \ioautomaton{}, we need at least as many states as the size of
  the example (plus one) to recognize it. Therefore, the minimal layered 
  \ioautomaton{} that recognizes one given input/output example is easy to
  construct, by using that many states.
\end{remark}


\section{Solving the Minimization Problem in NP}
\label{section:easiness}

We now focus on finding an algorithm for solving the minimization
problems~\ref{pb:sample} and \ref{pb:decsample}. In this section, we propose an
approach which solves the problem in non-deterministic polynomial-time. Combined
with the proofs in the previous sections, we can deduce that
Problem~\ref{pb:decsample} is $\NP$-complete.

The key is to prove (see Lemma~\ref{lemma:sampleio}, proof in
Appendix~\ref{app:sampleio}) that for any valid set of input/output examples,
there exists a total \ioautomaton, consistent with $E$, and whose size is at
most $2 + \sum_{w \in E} |w|$. Then, a naive minimization approach can iterate
through all integers $i$ between $1$ and this bound, guess non-deterministically
a DFA $A$ of size $i$, and check whether $A$ is a total
\ioautomaton{} consistent with $E$. We prove that this final check can be done
in polynomial time (see Lemma~\ref{lemma:ioautomaton}), meaning that the whole
procedure has non-deterministic polynomial time.

\begin{restatable}{lemma}{sampleio}
\label{lemma:sampleio} Let $E \subseteq (\Sigma \times \Gamma)^*$ be a valid set
of input/output examples. There exists a total \ioautomaton{}, consistent with
$E$, with at most $2 + \sum_{w \in E} |w|$ states.
\end{restatable}

Checking whether a DFA $A$ is a total \ioautomaton{} can be done in 
polynomial time, as shown in Lemma~\ref{lemma:ioautomaton}.
In addition, checking whether an \ioautomaton{} $A$ is consistent with $E$,
can be done by doing membership checks $w \in A$ for each $w \in E$.

\begin{lemma}
\label{lemma:ioautomaton}
Let $A$ be a DFA over the alphabet $\Sigma \times \Gamma$.
We can check in polynomial time whether $A$ is a total \ioautomaton{}.
\end{lemma}

\begin{proof} 
Let $A'$ be the projection of $A$ over the input part of the alphabet $\Sigma$.
The output-uniqueness property of $A$ is equivalent to the fact that $A'$ is
unambiguous. Checking whether an NFA is unambiguous can be done in polynomial
time~\cite{DBLP:books/daglib/0023547}.

For the output existence property, we check whether $\Sigma^* = A'$, which can
be done in polynomial time~\cite{DBLP:journals/siamcomp/StearnsH85} since $A'$
has been verified to be unambiguous.
\end{proof}

Using these lemmas, we conclude with the main result of this section.

\begin{theorem}
The minimization problems (\ref{pb:sample}, \ref{pb:decsample}, and 
\ref{pb:layered})
can be solved in $\NP$.
\end{theorem}

\section{Algorithm for Solving the Minimization Problem}
\label{section:algorithm}

\subsection{Description of the Algorithm}

The algorithm given in the previous section is not applicable in practice, as it
requires \emph{guessing} a total \ioautomaton{} that satisfies the constraints.
On a computer, this would require enumerating all automata of a certain size
until we find one that satisfies the constraints.

In this section, we instead propose to encode the constraints in a logical
formula, and let an SMT solver check satisfiability of the formula. More
precisely, given a set of input/output examples $E \subseteq (\Sigma \times
\Gamma)^*$, and $k \geq 1$, we define a formula $\varphi_{E,k}$ which is
satisfiable if and only if there exists a total 
\ioautomaton{} with $k$ states and that is consistent with $E$.

Then, in order to find the minimal total \ioautomaton{} with a given set 
of examples $E$, our algorithm checks satisfiability of 
$\varphi_{E,1}$, then $\varphi_{E,2}$, and so on, until 
one of the formula is satisfiable and the automaton is found. 

Encoding all the constraints of the problem in a logical formula is challenging.
The main reason is that SMT solver are best suited for dealing with logical
formula written in purely existential form, while the constraints that we want
to express (totality and output-uniqueness for \ioautomata{}) are naturally
expressed using alternations between \emph{for all} and \emph{exists} 
quantifiers. Still, we were able to find a purely existential encoding of the
problem, which we describe below. 

\subsection{Encoding}

\label{sec:encoding}

The free variables of $\varphi_{E,k}$ are functions that describe a DFA $A$ with
$k$ states. More precisely, $\varphi_{E,k}$ contains a free variable $\delta: Q
\times (\Sigma \times \Gamma) \rightarrow Q$ describing the transition relation
where $Q$ is a finite domain $\set{q_1,\dots,q_k}$. The formula $\varphi_{E,k}$
also contains a formula $\isFinal: Q \rightarrow \set{\bot,\top}$ specifying 
the final states. By convention, $q_1$ is the initial state, and $q_k$ is a 
non-accepting sink state. 

We also add a free variable $\deltain: Q \times \Sigma \times Q$ describing the
projection $A'$ of $A$ over the input alphabet $\Sigma$. The variable $\deltain$
is expressed as a relation rather than as a function, since in general, $A'$ can
be non-deterministic.

The formula $\varphi_{E,k}$ is then composed of multiple components:
\[
  \acceptexamples \land \corresp \land \isunambiguous \land \istotal.
\]

The formula $\acceptexamples$ constrains the transition relation $\delta$ and
the accepting states $\isFinal$ to make sure that every input/output example in
$E$ is accepted by $A$. The formula $\corresp$ ensures that the variable
$\deltain$ indeed represents the projection of $\delta$ on the input alphabet
$\Sigma$.

The formulas $\isunambiguous$ and $\istotal$ correspond to the approach
described in Lemma~\ref{lemma:ioautomaton}. The formula $\isunambiguous$ is a
constraint over the variables $\deltain$ and $\isFinal$, representing the
projection $A'$. It states that $A'$ is a UFA, which ensures that $A$ is an
\ioautomaton{}. Being unambiguous is naturally stated using quantifiers: for
every word $w$, if $w$ is accepted by two runs $r_1$ and $r_2$ in $A'$, then
$r_1$ and $r_2$ must be identical runs (i.e.~going through identical states).
However, writing this condition as is would make it hard for the SMT solver to
check satisfiability of the formula, due to the universal quantification.

Instead, our formula $\isunambiguous$ is inspired from the algorithm that checks
whether a given NFA is unambiguous~\cite{DBLP:books/daglib/0023547}. This
algorithm constructs inductively the pairs of states $(q_i,q_j)$ that are
reachable by the same word, but with distinct runs. Then, the NFA is unambiguous
if and only if there exists a pair $(q,q')$ in that inductive construction where
$q$ and $q'$ are both final states. 

The construction starts with the empty set, and adds, for each state $q$ which
is reachable, and for every letter $a \in \Sigma$, the pairs $(q_1,q_2)$, with
$q_1 \neq q_2$ such that $\deltain(q,a,q_1)$ and $\deltain(q,a,q_2)$ hold. Then,
for every $(q_i,q_j)$ and every $a \in \Sigma$, we add the pairs $(q_i',q_j')$
such that $\deltain(q_i,a,q_i')$ and $\deltain(q_j,a,q_j')$ hold.

Therefore, to ensure the unambiguity $A'$, the formula $\isunambiguous$ states
that there exists a fixed point (a set of pairs of states) to that construction,
i.e.~a set which is closed under adding new pairs according to the rules above,
and which does not contain a pair $(q,q')$ where $\isFinal(q)$ and 
$\isFinal(q')$ hold. 

The formula $\istotal$ is also a constraint over the variables $\deltain$ and
$\isFinal$, and states that $A'$ recognizes every string in $\Sigma^*$. This
ensures that the \ioautomaton{} $A$ is total. Again, this constraint is
naturally expressed using quantifiers: for every word $w$, there exists a run
for $w$ in $A'$. Such formulas are challenging for SMT solvers. Instead,
our formula relies on the fact that $A'$ is ensured to be unambiguous by the
formula $\isunambiguous$. More precisely, to check that $A'$ accepts every
string of $\Sigma^*$, it suffices to check that $A'$ has $|\Sigma|^l$ accepting
runs, for every $l \geq 0$. Moreover, it can be shown that it is enough
(see~\cite{DBLP:journals/siamcomp/StearnsH85}) to do this check so for $l \leq
|Q|$. 
 
Our formula $\istotal$ introduces free variables $c_{l,q}$, for each $0 \leq l
\leq |Q|$, and $q \in Q$, and constrains them so that they count how many runs
of length $l$ end in state $q$. $\istotal$ then states that for every $0 \leq l
\leq |Q|$, the number of accepting runs of length $l$ equals $|Sigma|^l$, i.e.
\(
  \sum_{q \in Q} \booltoint{\isFinal(q)} * c_{l,q} = |\Sigma|^l
\) 
where $\booltoint{\top} = 1$, and $\booltoint{\bot} = 0$.

\section{Experimental Evaluation}
\label{section:experiments}

We implemented our algorithm in Scala, using Z3~\cite{DBLP:conf/tacas/MouraB08}
as our backend. 


\subsection{Discovering Small Automata for Common Functions}

We give in this section a few examples that we ran using our algorithm. We focus
on examples that have small automata, whether or not the input examples are
small. Indeed, the combinatorial explosion makes it hard for the SMT solver to
find solutions for automata that have more than ~$10$ states. The results are
shown in Figure~\ref{table:examples}. The examples operate on binary 
representations of numbers, truncating the output to the length of inputs where
needed. We note that simple relations such as addition are recovered from
examples without the need to specify any expression grammars as in Syntax-Guided
Synthesis \cite{alur2013}, because automaton minimality provides the needed bias
towards simple solutions. Adding more examples than needed (e.g. 22 examples of
length 22) keeps the synthesis time manageable, which is useful for cases of
automatically generated examples.

\begin{figure}[b]
\begin{tabular}{c|c|@{\ }r@{\ \ }|c|c|c}
Problem & \#Examples & Ex.\ Length & \#Aut.\ States & Alphabet S. & Time (sec.) \\\hline
$x,y \mapsto x{+}y$ &  1     & 17    & 3      & 8           & 0.40 \\
$x,y \mapsto x{+}y$ &  5     & 4      & 2     & 8        & 0.37\\
$x,y \mapsto x{+}y$ &  22     & 22      & 2   & 8          & 0.60\\
xor       & 1     & 4      & 2        & 8    & 0.11\\
and       & 1     & 4      & 2        & 8    & 0.13\\
or       & 1      &  4     & 2        & 8    & 0.13\\
not       & 1      &  4     & 2       & 4     & 0.18\\
$x \mapsto 2x+1$ & 1 & 5    & 3       & 4     & 0.35 \\ 
$(p \lor q) \land (r \lor s) \land \neg t$ & 1 & 32 & 2 & 64 & 0.28\\
$(p \lor q) \land (r \lor s) \land \neg t$ & 32 & 1 & 2 & 64 & 0.36\\
\end{tabular}
\caption{Synthesis of some common functions from examples, showing
successful discovery of minimal automata and tolerance to many long examples
and larger alphabets.
\label{table:examples}}
\end{figure}
 
\subsection{Evaluating Usefulness of Minimality on Random Automata}

The next set of experiments evaluate the likelihood that our
algorithm finds the automaton that the user has is mind, depending
on the number and size of the input/output examples provided. We generated $100$
random minimal total \ioautomata{} with $5$ states, where the input and output
alphabet were both of size $2$. For each \ioautomaton{} $A$, and for every $1
\leq i,j \leq 15$, we generated $i$ random words in $\Sigma^*$, of length $j$.
For each such word, we looked up the corresponding output in $A$, thereby
constructing a set of input/output examples $E$ for $A$. 
Then, we used our
algorithm on $E$ to see whether the obtained automaton would be
$A$. In Table~\ref{table:experiments} (in Appendix~\ref{app:tables}), we
summarized, for every $i$ and $j$, out of the $100$ automata, how many we were
able to reobtain using that method. Overall, the experiments ran for about $3$
hours, for $15*15*100 = 22500$ queries. The $3$ hours also include
the time taken to generate the random automata.
To generate a random minimal total \ioautomaton{}, we generated a random sample,
and applied our algorithm. Then, if the obtained automaton had $5$ states, we
kept it for our experiment. Our selection for the choice of the random automata
is therefore biased, as the automata are found by our tool in the first place.

\paragraph{Discussion.}
Generally, the results show that the greater the number of examples given, and 
the longer they are, the more likely we are to find the automaton that we want.
More interestingly, we note that we are more likely to find the automaton we 
want with a large number of small examples (e.g.~$i = 15, j = 5$) than with a 
small number of large examples (e.g.~$i = 5, j = 15$). 

Another interesting observation is that the likelihood of finding the automaton
increases sharply when using examples of size $j = 4$ rather than $j = 3$.
Without counting the sink state, the automata we considered have $4$ states.
This suggests that in general, a good strategy is to give multiple examples
which are at most as long as the number of states (though the user giving the
examples may not know how many states are required for the minimal automaton).



\section{Related Work}

In~\cite{DBLP:conf/ecoop/MayerHK17}, we studied the problem of synthesizing
tree-to-string transducers from examples. Here, instead of having the user
provide input/output examples, we proposed an algorithm that generates
particular inputs, and asks the user what are the corresponding outputs. We show
that, when the algorithm is allowed to analyze previous answers in order to
generate the next question, then the number of questions required to determine
the transducer that the user has in mind is greatly reduced (compared to an
approach without interaction, where the algorithm would ask for all outputs at
once).

The results obtained in~\cite{DBLP:conf/ecoop/MayerHK17} do not directly apply
here, as they were for single-state transducers. However, some of the techniques
are fundamental and could be reused here. In that respect, we could generate
questions for the users, and guarantee that the generated \ioautomaton{} is
indeed the one that the user had in mind (given some bound on the number of
states). 

Our paper is similar in spirit to \cite{DBLP:journals/iandc/Gold78}, where
the author proves that Problem~\ref{pb:decsample} is $\NP$-complete for
\emph{deterministic} Mealy machines. Their $\NP$-hardness holds even when the
alphabets' sizes are fixed to $2$, but the case where the number of states is
fixed is not treated. Moroever, even though \ioautomata{} are a more general
model than deterministic Mealy machines, the $\NP$-hardness
of~\cite{DBLP:journals/iandc/Gold78} cannot be directly applied to
\ioautomata{}. 

There is a long line of work devoted to learning \emph{deterministic} finite
state transducers (see
e.g.~\cite{DBLP:conf/icalp/Bojanczyk14,DBLP:journals/pami/OncinaGV93,DBLP:journals/ml/AartsKTVV14,merten2013active}).
Algorithms for learning deterministic finite automata~(e.g.~\cite{angluin1987})
or finite transducers do not directly translate to our setting, since we need to
consider functionality and totality constraints, as shown in
Section~\ref{sec:encoding}.
Methods for learning non-deterministic automata~(e.g.~\cite{DBLP:conf/ijcai/BolligHKL09})
do not directly apply to our setting either, for the same reasons.

A particular case of learning transducers is an interpolation problem, that
consists in learning a finite automaton that accepts some given inputs
(i.e.~outputs $1$) and rejects some other inputs (i.e.~outputs $0$)~(see
e.g.~\cite{DBLP:journals/jacm/PittW93,DBLP:conf/tacas/ChenFCTW09,
DBLP:conf/cade/GrinchteinLP06}).

In~\cite{DBLP:conf/icgi/KhaliliT14}, the authors present an algorithm for
learning \emph{non-deterministic} Mealy machines. They are interested in
non-determinism to represent unknown components of reactive systems, and as such
do not focus on \emph{functional} non-deterministic Mealy machines. Moreover,
their focus is rather on the algorithmic aspect of the problem rather than on
complexity classes.

In~\cite{DBLP:conf/popl/Gulwani11}, the author proposes an efficient synthesis
procedure from examples for a language that does string transformations, but
does not deal with the issue of synthesizing finite-state transducers.
Our algorithm in Section~\ref{section:algorithm} is inspired from the bounded
synthesis approach of~\cite{DBLP:journals/sttt/FinkbeinerS13}. There, the
authors suggest that bounding the number of states is a good strategy to
synthesize reactive systems. They also propose a reduction from the bounded 
synthesis problem for reactive systems to SMT solvers.

In~\cite{DBLP:conf/fmcad/HamzaJK10}, we presented a way to synthesize
string-to-string functions given any specification written in weak monadic
second-order logic. Using these techniques, it would be possible to synthesize
an \ioautomaton{} consistent with input/output examples, by writing the
input/output examples as a logical formula. However, this approach would not
yield the minimal \ioautomaton{} consistent with the examples. For example,
regardless of how many input/output examples we give for the function
$(\set{0,1} \times \set{0,1})^* \rightarrow \set{0,1}^*$ which xor's two streams
of bits, this approach would not yield the $1$-state automaton that we are
expecting. Instead, the method will generate large automata that are consistent
with the given examples, but do not recognize the xor operation for other input
strings. On the other hand, our approach can find this automaton with only a few
small examples.


The automata we consider in this paper are closely related to the notion of
\emph{thin language} (see e.g.~\cite{DBLP:journals/dam/PaunS95}). A language $L$
is called thin if for every $n \in \Nat$, it contains at most one word of length
$n$. Moreover, $L$ is called length-complete if for every $n \in \Nat$, $L$
contains at least one word of length $n$.
When $|\Sigma| = 1$, i.e.~when only the length of the input matters, our
minimization problem corresponds exactly to finding a minimal DFA that contains
a given set of examples, which is both thin and length-complete. We left this
question open in Section~\ref{sec:summary}, and leave it for future work.
This analogy with thin languages breaks when using a non-unary input alphabet.

In \cite{DBLP:conf/lata/SmetsersFV18}, the authors encode the problem of
learning DFAs in an SMT solver. As is the case with our algorithm, such
encodings only perform well for finding automata with a small number of states
(up to $10$ or $15$).

\section{Conclusions}

\ioautomata{} are a form of functional non-deterministic one-way finite-state
transducers (see e.g.~\cite{DBLP:books/daglib/0023547,DBLP:books/lib/Berstel79})
where each transition is forced to produce exactly one letter (instead of $0$ or
more in the general case). The term functional corresponds to the output uniqueness
property of \ioautomata{}, and ensures that despite the non-determinism, at most
one output string is produced for each input string. The non-determinism here
refers to the input part of the alphabet, and \ioautomata{}, even though they
are deterministic on $\Sigma \times \Gamma$, can indeed be non-deterministic in
the input alphabet $\Sigma$.
In that sense, \ioautomata{} can define transformations that are not captured by
deterministic one-way transducers, such as the function that maps a word 
$w$ to $l^{|w|}$ where $l$ is the last letter of $w$. On the other hand, 
deterministic one-way transducers can recognize transformations not 
recognized by \ioautomata, since they do not require the output to have the 
same length as the input. This can be circumvented
by padding the input and output strings using a dummy letter.
Existing synthesis algorithms generally target classes of deterministic
transducers, such as subsequential transducers (see
e.g.~\cite{DBLP:conf/icgi/Vilar96}). Our results about \ioautomata{} are a first
step towards synthesis algorithm for larger classes of deterministic or
functional non-deterministic transducers, such as two-way finite-state
transducers, or streaming string transducers~\cite{alur2010}. We have shown that
most variants of synthesis for \ioautomata{} are $\NP$-complete, and presented a promising
approach using an encoding into SMT formulas.

\section{Acknowledgement}

We would like to thank the anonymous reviewers for their thorough comments and
for relevant references related to learning finite automata and regarding the
interpolation problem. 

\newpage
\bibliographystyle{eptcs} 
\bibliography{references}

\newpage
\appendix


\section{Proof of $\NP$-Hardness of the Minimization Problem with 
One Input/Output Example and Fixed Alphabets}
\label{app:fixedalphabets}

\fixedalphabets*

\begin{proof}
  Consider an instance $\varphi$ of \oneinthree{}, with a set of variables 
  $V = \set{x_0,\dots,x_{m-1}}$ with $m \geq 1$, 
  and a set of clauses $C \subseteq V^3$.
  Without loss of generality, we assume that $(x_i,x_j,x_k) \in C$
  implies $i < j < k$ (the variables that appear in a clause are ordered).
  We reduce \oneinthree{} to Problem~\ref{pb:decsample} as follows.
  We define $\Sigma = \set{a,b,c,d}$ and $\Gamma = \set{0,1}$ (fixed alphabets),
  and $n = 3m$.
  
  Then, we define $E = \set{w}$ where $w$ is the input/output examples made of 
  the concatenation of the following words 
  (the word $a^{2n} \zip 0^{n-1}10^{n-1}1$ must go first in the concatenation,
  but the other words can be concatenated in any order):
  \begin{itemize}
  \item $a^{2n} \zip 0^{n-1}10^{n-1}1$, \vspace{1ex}
  \item $b^i a^{n-i} \zip 0^{n-1}1$ for $1 \leq i < m$,
  \item $a^m b^i a^{2m-i} \zip 0^{n-1}1$ for $1 \leq i < m$,
  \item $a^{2m} b^i a^{m-i} \zip 0^{n-1}1$ for $1 \leq i < m$, \vspace{1ex}
  \item $b^m a^{n} \zip 0^{n+m-1}1$,
  \item $a^m b^m a^n \zip 0^{n+2m-1}1$,
  \item $a^{2m} b^m a^n \zip 0^{2n-1}1$, \vspace{1ex}
  \item $a^i c a^n \zip 0^{n+i} 1$ for $0 \leq i < 2m$, 
  \item $a^i c a^n \zip 0^i 1 0^{n-1} 1$ for $2m \leq i < 3m$, \vspace{1ex}
  \item $a^i d b^{m-i} a^n \zip 0^{n+m}1$ for $0 \leq i < m$,
  \item $a^{m+i} d b^{m-i} a^n \zip 0^{n+2m} 1$ for $0 \leq i < m$,
  \item $a^{2m+i} d b^{m-i} a^n \zip 0^{2n} 1$ for $0 \leq i < m$,  \vspace{1ex}
  \item $a^i d c a^n \zip 0^{n+1}1$ for $0 \leq i < m$, \vspace{1ex}
  \item $a^i ddd a^{n-i} \zip 0^{n+2}1$ for $0 \leq i < n$,  \vspace{1ex}
  \item $a^i d a^{j-i} d a^{k-j} d a^{2m-k} \zip 0^{2m+2}1$ for $(x_i,x_j,x_k) \in C$.
  \end{itemize}
  
  We prove that $\varphi$ has a satisfying assignment if and only if 
  there exists a total \ioautomaton{} $A$, consistent with $E$, and with 
  (at most) $n$ states.
  
  $(\Rightarrow)$
  Let $f: V \rightarrow \set{\bot,\top}$ be a satisfying assignment for 
  $\varphi$.
  We define $A = (\Sigma\times\Gamma,Q,\initState,\delta,F)$ following 
  Figure~\ref{fig:base2} with $Q = F = \set{q_0,q_1,\dots,q_{3m-1}}$ and 
  $\initState = q_0$.
  
  Then, for each $x_i \in V$ for $0 \leq i < m$, if $f(x_i) = \top$, we add 
  three transitions in $\delta$, called \emph{forward transitions}:
  $(q_i, (d,0), q_{m+i})$, 
  $(q_{m+i}, (d,0), q_{2m+i})$, and
  $(q_{2m+i}, (d,0), q_i)$. 
  If $f(x_i) = \bot$, we add three transitions as well, called 
  \emph{looping transitions}:
  $(q_i, (d,0), q_i)$, 
  $(q_{m+i}, (d,0), q_{m+i})$, and
  $(q_{2m+i}, (d,0), q_{2m+i})$. 
  Note that the definitions of forward and looping transitions are similar, 
  but different from the notions introduced in the proof of 
  Theorem~\ref{th:fixedstates-oneexample}.
  
  $A$ is a total \ioautomaton{}, since all states are final, and for every 
  state and every input in $\Sigma$, there is a unique outgoing transition
  labeled by this input (and some output in $\Gamma$). Moreover, we can verify
  that $A$ is consistent with all the input/output example $w$.
  
  $(\Leftarrow)$
  Let $A = (\Sigma\times\Gamma,Q,\initState,\delta,F)$ be a total \ioautomaton{}
  consistent with $E$ and with $n$ states.
  Without loss of generality, 
  we let $Q = \set{q_0,q_1,\dots,q_{n-1}}$ with $\initState = q_0$.
  
  The overall idea of the proof is similar to the one of 
  Theorem~\ref{th:fixedstates-oneexample}.
  First, using Lemma~\ref{lemma:a01loop} and Lemma~\ref{lemma:splitexamples},
  we deduce that $A$ must have an $(a,0,1)$-loop, and must accept all the 
  individual words that constitute the concatenation $w$.
  
  Second, we prove that the transitions involving $b,c \in \Sigma$ must be
  as described by Figure~\ref{fig:base}.
  This is due to the examples:
  \begin{itemize}
  \item $b^i a^{n-i} \zip 0^{n-1}1$ for $1 \leq i < m$,
  \item $a^m b^i a^{2m-i} \zip 0^{n-1}1$ for $1 \leq i < m$,
  \item $a^{2m} b^i a^{m-i} \zip 0^{n-1}1$ for $1 \leq i < m$, \vspace{1ex}
  \item $b^m a^{n} \zip 0^{n+m-1}1$,
  \item $a^m b^m a^n \zip 0^{n+2m-1}1$,
  \item $a^{2m} b^m a^n \zip 0^{2n-1}1$, \vspace{1ex}
  \item $a^i c a^n \zip 0^{n+i} 1$ for $0 \leq i < 2m$, 
  \item $a^i c a^n \zip 0^i 1 0^{n-1} 1$ for $2m \leq i < 3m$.
  \end{itemize}
  
  Third, we prove that each variable $x \in V$ must have either 
  forward transitions, or looping transitions. We prove this using 
  the examples:
  \begin{itemize}
  \item $a^i d b^{m-i} a^n \zip 0^{n+m}1$ for $0 \leq i < m$,
  \item $a^{m+i} d b^{m-i} a^n \zip 0^{n+2m} 1$ for $0 \leq i < m$,
  \item $a^{2m+i} d b^{m-i} a^n \zip 0^{2n} 1$ for $0 \leq i < m$,  \vspace{1ex}
  \item $a^i d c a^n \zip 0^{n+1}1$ for $0 \leq i < m$, \vspace{1ex}
  \item $a^i ddd a^{n-i} \zip 0^{n+2}1$ for $0 \leq i < n$.
  \end{itemize}
  We then use this fact to define an assignment $f$ that assigns 
  $\top$ to variables that have forward transitions, and $\bot$ to variables 
  that have looping transitions.
  
  Finally, for each clause $(x_i,x_j,x_k) \in C$, using that 
  $A(a^i d a^{j-i} d a^{k-j} d a^{2m-k}) = 0^{2m+2}1$, 
  we deduce that exactly one variable out of 
  $x_i$, $x_j$ and $x_k$ must have forward transitions, and conclude that 
  $f$ is a satisfying assignment for $\varphi$.
  
  We now give more details for each step of the proof.
  Our first goal is to prove that the transitions involving $b,c \in \Sigma$ are 
  as described in Figure~\ref{fig:base2}.
  Consider the input/output examples
  $b^i a^{n-i} \zip 0^{n-1}1 \in A$ for $1 \leq i < m$.
  We deduce that when reading $b^i \zip 0^i$ from the initial state,
  $A$ must end in state $q_{i+1}$.
  This implies that for $0 \leq i < m-1$, there is a transition 
  $(q_i,(b,0),q_{i+1}) \in \delta$.
  
  Using the same reasoning, 
  we deduce from $b^m a^{n} \zip 0^{n+m-1}1 \in A$ that there is 
  a transition $(q_{m-1},(b,1),q_0) \in \delta$.
  Similarly, we can prove that the transitions with letter $b \in \Sigma$ from 
  the states $q_m,q_{m+1},\dots,q_{3m-1}$ are as described in 
  Figure~\ref{fig:base2}.

  The input/output examples 
  $A(a^i c a^n) = 0^{n+i} 1$ for $0 \leq i < 2m$, and 
  $A(a^i c a^n) = 0^i 1 0^{n-1} 1$ for $2m \leq i < 3m$,
  imply that the transitions ensure that from every state, there is a
  transition labeled by $c$ going to $q_0$.
  From the first two columns (states $q_0$ to $q_{2m-1}$), 
  these transitions are labeled by $(c,0)$,
  while from the last column (states $q_{2m}$ to $q_{3m-1}$),
  these transitions are labeled by $(c,1)$.
  This is as depicted in Figure~\ref{fig:base2}.

  Then, we want to prove that for each $x_i \in V$, $0 \leq i < m$, 
  $x_i$ either has forward transitions or looping transitions.
  Let $i \in \set{0,\dots,m-1}$.
  Consider the fact that $a^i d b^{m-i} a^n \zip 0^{n+m}1 \in A$.
  After reading $a^i \zip 0^i$, the automaton $A$ is in state $q_i$.
  Moreover, the only states from which $b^{m-i} a^n \zip 0^{n+m-i-1}1$ is 
  accepted are $q_i$, $q_{m+i}$ and $q_{2m+i}$. 
  Therefore, there should either exist a transition 
  $(q_i,(d,0),q_i) \in \delta$, or
  $(q_i,(d,0),q_{m+i}) \in \delta$, or
  $(q_i,(d,0),q_{2m+i}) \in \delta$.
  The last option is not possible, due to the input/output example
  $a^i d c a^n \zip 0^{n+1}1 \in A$.
  
  There are then two cases to consider: 
  $(q_i,(d,0),q_i) \in \delta$, or
  $(q_i,(d,0),q_{m+i}) \in \delta$.
  When $(q_i,(d,0),q_{m+i}) \in \delta$, we prove that $x_i$ has forward 
  transitions.   
  Consider the example $a^i ddd a^{n-i} \zip 0^{n+2}1 \in A$.
  After reading $a^i \zip 0^i$, the automaton $A$ must be in state $q_i$.
  Moreover, the only state from which $A$ can accept 
  $a^{n-i} \zip 0^{n-i-1}1$ is $q_i$ as well.
  Therefore, $A$ must go from $q_i$ to $q_i$ when reading 
  $ddd \zip 000$. Since $(q_i,(d,0),q_{m+i}) \in \delta$,
  $A$ must go from $q_{m+i}$ to $q_i$ when reading $dd \zip 00$.
  
  The constraint $a^{m+i} d b^{m-i} a^n \zip 0^{n+2m} 1$
  enforces the existence of an outgoing transition, labeled by $(d,0)$, 
  from $q_{m+i}$ to either $q_i$ or $q_{m+i}$ or $q_{2m+i}$ 
  as these three states are the only states from where 
  $b^{m-i} a^n \zip 0^{n+m-i-1}1$ is accepted. 
  Due to this constraint, the only possibility
  for $A$ to go from $q_{m+i}$ to $q_i$ when reading $dd \zip 00$ is
  to have a transition
  $(q_{m+i},(d,0),q_{2m+i}) \in \delta$ and then an additional transition
  $(q_{2m+i},(d,0),q_{i}) \in \delta$.
  This proves that $x_i$ has forward transitions.
  
  The other case to consider is when $(q_i,(d,0),q_i) \in \delta$.
  Here we prove that $x_i$ has looping transitions.
  Using the fact that 
  $a^{m+i} ddd a^{n-m-i} \zip 0^{n+2}1 \in A$, we know that
  $A$ must go from $q_{m+i}$ to $q_{m+i}$ when reading 
  $dd \zip 000$.
  Then, using the same reasoning as above 
  with the examples 
  $a^{m+i} d b^{m-i} \zip 0^m 1 \in A$ and 
  $a^{2m+i} d b^{m-i} \zip 0^m 1 \in A$, we deduce
  that there must exist transitions, labeled by $(d,0)$ 
  from $q_{m+i}$ and $q_{2m+i}$ to either 
  $q_i$, $q_{m+i}$ or $q_{2m+i}$.
  Combined with the fact that $(q_i,(d,0),q_i) \in \delta$, the 
  only possibility for $A$ to go from $q_{m+i}$ to $q_{m+i}$ when reading 
  $ddd \zip 000$ is to have a 
  transition $(q_{m+i},(d,0),q_{m+i}) \in \delta$.
  Similarly, using 
  $a^{2m+i} ddd a^{n-2m-i} \zip 0^{n+2}1 \in A$, we deduce 
  that there is a
  transition $(q_{2m+i},(d,0),q_{2m+i}) \in \delta$.
  We have proved that, in that case, $x_i$ has looping transitions.
  
  Overall, we have proved that $x_i$ either has forward transitions, 
  or looping transitions. We define the assignment $f$ that assigns 
  $\top$ to variables that have forward transitions, and $\bot$ to variables 
  that have looping transitions.
  Let $(x_i,x_j,x_k) \in C$. 
  We know $a^i d a^{j-i} d a^{k-j} d a^{2m-k} \zip 0^{2m+2}1 \in A$.
  The only state from which $a^{2m-k} \zip 0^{2m-k-1}1$ can be accepted 
  is $q_{m+k}$.
  Therefore, $A$ must end in state $q_{m+k}$ after reading 
  $a^i c a^{j-i} c a^{k-j} c \zip 0^{k+3}$.
  For this, exactly one of the variables $x_i$, $x_j$ and $x_k$ must have
  forward transitions, while the two others must have looping transitions.
  This concludes the proof that $f$ is a satisfying assignment for $\varphi$.
  
  \end{proof}
  

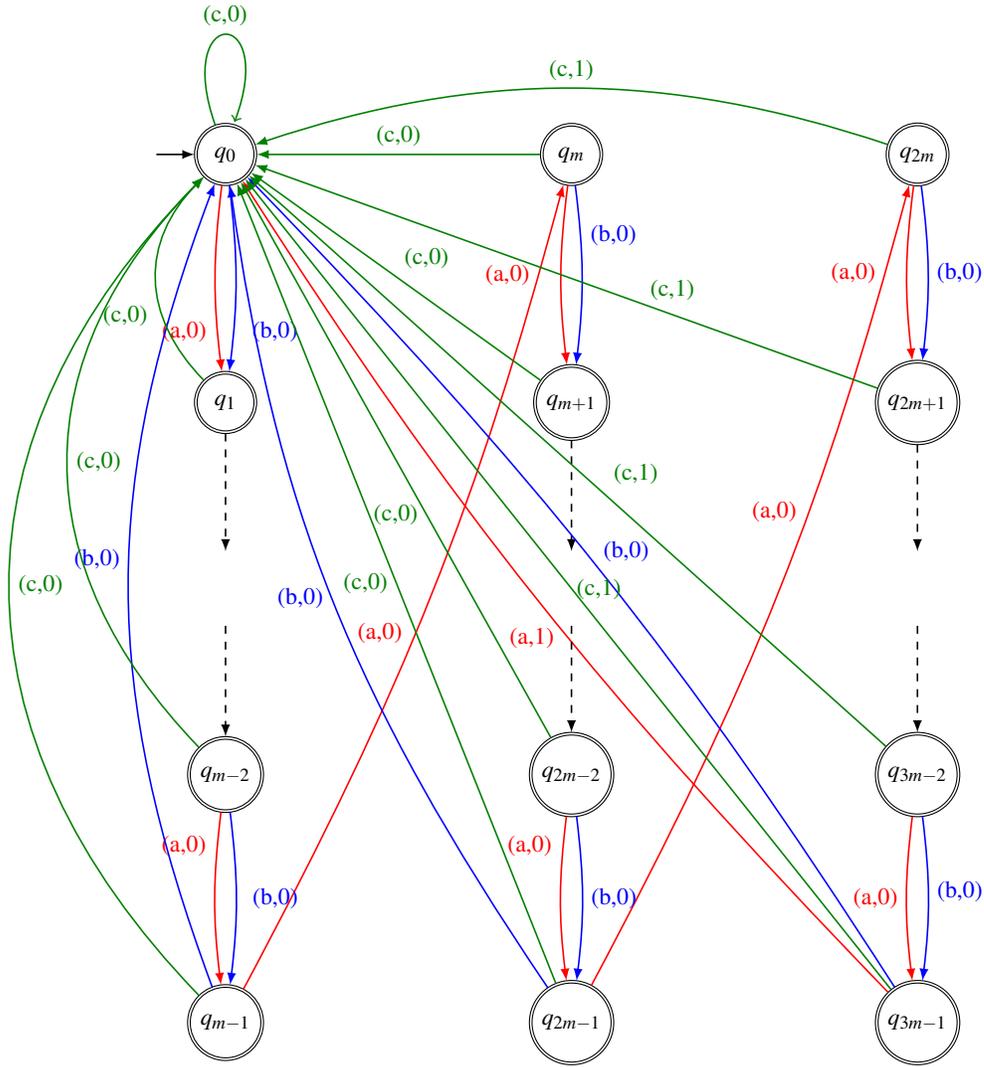
\begin{figure}
\centering

\hspace{-5em}
\begin{tikzpicture}[xscale=4.6,yscale=3.3]

\footnotesize

\layer{0}{q_0}{q_m}{q_{2m}}{0}
\layer{-1}{q_1}{q_{m+1}}{q_{2m+1}}{1}
\layer{-2.5}{q_{m-2}}{q_{2m-2}}{q_{3m-2}}{2}
\layer{-3.5}{q_{m-1}}{q_{2m-1}}{q_{3m-1}}{3}

\draw[tr] (-0.2,0) -- (a0);


\path[tra] (a3) edge[bend right=5] node[below left=4mm] { (a,0) } (b0);
\path[tra] (b3) edge[bend right=5] node[left,yshift=30,xshift=10] { (a,0) } (c0);
\path[tra] (c3) edge[bend left=5] node[left,yshift=-15,xshift=10] { (a,1) } (a0);


\path[tra] (a0) 
  edge[bend right=5] 
  node[left,yshift=-20] 
  { (a,0) } 
  (a1);
\path[tra] (b0) 
  edge[bend right=5] 
  node[left=3mm]
  { (a,0) }
  (b1);
\path[tra] (c0) 
  edge[bend right=5] 
  node[left=3mm] 
  { (a,0) } 
  (c1);

  
\path[trb] (a0) 
  edge[bend left=5] 
  node[right=1mm, yshift=-20] 
  { (b,0) } 
  (a1);
\path[trb] (b0) 
  edge[bend left=5] 
  node[right,yshift=15]
  { (b,0) }
  (b1);
\path[trb] (c0) 
  edge[bend left=5] 
  node[right,text width=1cm] 
  { (b,0) } 
  (c1);

\path[tr,dashed] (a1) edge ($(a1)-(0,0.6)$);
\path[tr,dashed] (b1) edge ($(b1)-(0,0.6)$);
\path[tr,dashed] (c1) edge ($(c1)-(0,0.6)$);

\path[tr,dashed] ($(a2)+(0,0.6)$) edge (a2);
\path[tr,dashed] ($(b2)+(0,0.6)$) edge (b2);
\path[tr,dashed] ($(c2)+(0,0.6)$) edge (c2);


\path[tra] (a2) 
  edge[bend right=5] 
  node[left,yshift=20] 
  { (a,0) } 
  (a3);
\path[tra] (b2) 
  edge[bend right=5] 
  node[left,yshift=20]
  { (a,0) }
  (b3);
\path[tra] (c2) 
  edge[bend right=5] 
  node[left] 
  { (a,0) } 
  (c3);
  

\path[trb] (a2) 
  edge[bend left=5] 
  node[right=1mm, text width=1cm] 
  { (b,0) } 
  (a3);
\path[trb] (b2) 
  edge[bend left=5] 
  node[right, text width=1cm]
  { (b,0) }
  (b3);
\path[trb] (c2) 
  edge[bend left=5] 
  node[above=1mm,right, text width=1cm] 
  { (b,0) } 
  (c3);


\path[trb] (a3) 
    edge[bend left=15]
    node[left,yshift=10] { (b,0) }
    (a0);
\path[trb] (b3) 
    edge[bend left=10]
    node[left] { (b,0) }
    (a0);
\path[trb] (c3) 
    edge[bend right=5]
    node[above,xshift=10] { (b,0) }
    (a0);
  
  
\path[trc] (a0)
  edge[loop above]
  node[above]
  { (c,0) }
  (a0);
  
\path[trc] (a1)
  edge[bend left=35]
  node[below=5mm,left]
  { (c,0) }
  (a0);
  
\path[trc] (a2)
  edge[bend left=35]
  node[right]
  { (c,0) }
  (a0);
  
\path[trc] (a3)
  edge[bend left=35]
  node[right]
  { (c,0) }
  (a0);
  
  
\path[trc] (b0)
  edge[bend left=0]
  node[above]
  { (c,0) }
  (a0);
  
\path[trc] (b1)
  edge[bend left=0]
  node[above right]
  { (c,0) }
  (a0);
  
\path[trc] (b2)
  edge[bend left=0]
  node[below=3mm,yshift=-5]
  { (c,0) }
  (a0);
  
\path[trc] (b3)
  edge[bend left=0]
  node[left]
  { (c,0) }
  (a0);
  

\path[trc] (c0)
  edge[bend right=25]
  node[above]
  { (c,1) }
  (a0);
  
\path[trc] (c1)
  edge[bend left=0]
  node[right=10mm,yshift=-5]
  { (c,1) }
  (a0);
  
\path[trc] (c2)
  edge[bend left=0]
  node[right=5mm,yshift=-5]
  { (c,1) }
  (a0);
  
\path[trc] (c3)
  edge[bend left=0]
  node[above right,yshift=-9]
  { (c,1) }
  (a0);

\end{tikzpicture}

\caption{
    \ioautomaton{} used in the proof of Theorem~\ref{th:fixedalphabet-oneexample}.
}
\label{fig:base2}
\end{figure} 

\section{Proof of $\NP$-Hardness of the Minimization Problem for Layered
Automata}
\label{app:layered}


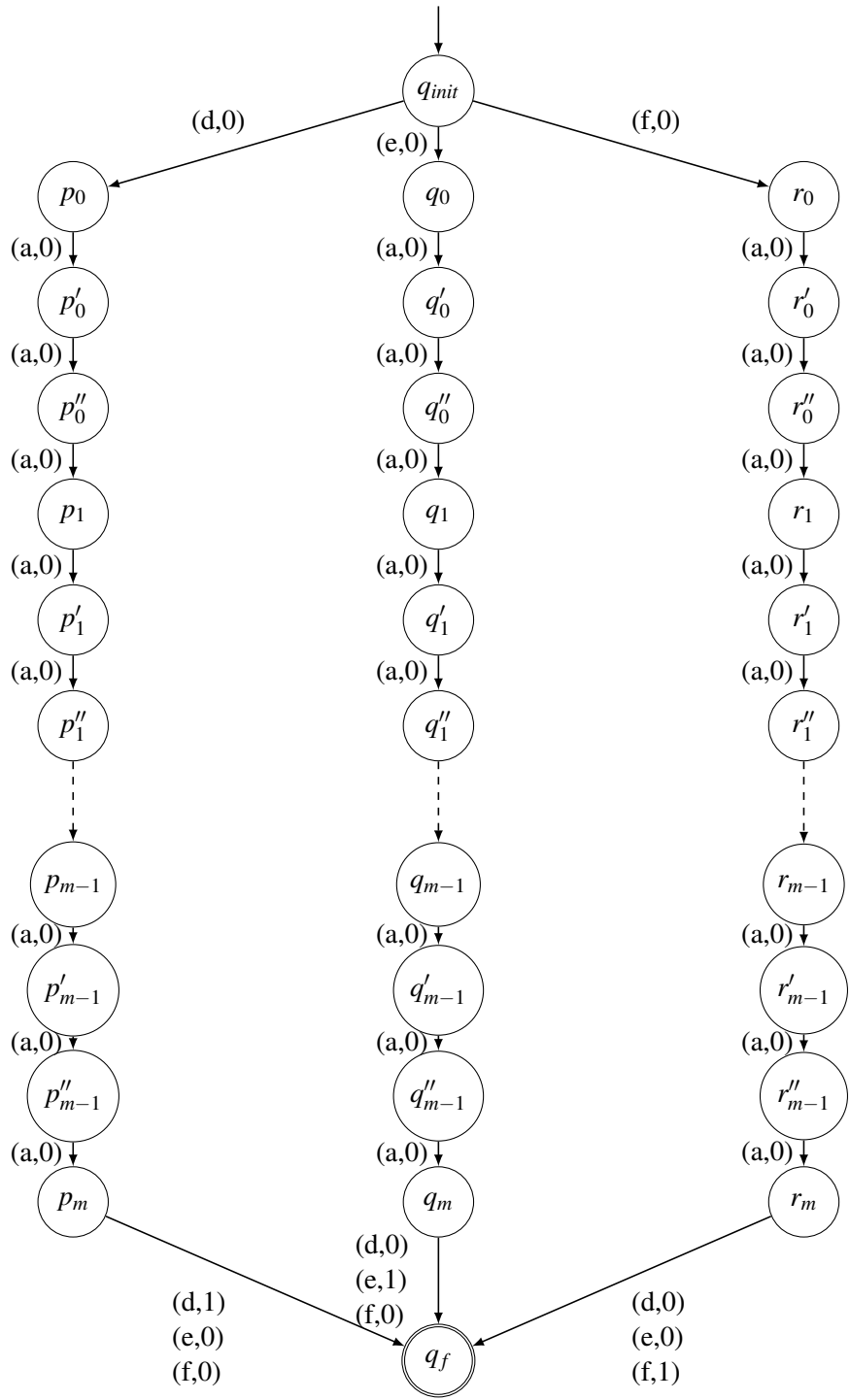
\begin{figure}
\centering

\begin{tikzpicture}[xscale=5,yscale=1.45]

\node[state] (h) at (1,1) {$\initState$};

\layern{0}{p_0}{q_0}{r_0}{0}
\layern{-1}{p_0'}{q_0'}{r_0'}{1}
\layern{-2}{p_0''}{q_0''}{r_0''}{2}

\layern{-3}{p_1}{q_1}{r_1}{3}
\layern{-4}{p_1'}{q_1'}{r_1'}{4}
\layern{-5}{p_1''}{q_1''}{r_1''}{5}

\layern{-6.5}{p_{m-1}}{q_{m-1}}{r_{m-1}}{6}
\layern{-7.5}{p_{m-1}'}{q_{m-1}'}{r_{m-1}'}{7}
\layern{-8.5}{p_{m-1}''}{q_{m-1}''}{r_{m-1}''}{8}
\layern{-9.5}{p_m}{q_m}{r_m}{9}

\node[state,accepting] (f) at (1,-11) { $q_f$ };

\draw[tr] (1,1.8) -- (h);

\path[tr] (h) edge node[above left] { (d,0) } (a0);
\path[tr] (h) edge node[left] { (e,0) } (b0);
\path[tr] (h) edge node[above right] { (f,0) } (c0);

\path[tr] (a9) edge node[below left, text width=1cm] 
  { (d,1) (e,0) (f,0) } (f);
\path[tr] (b9) edge node[left, text width=1cm] 
  { (d,0) (e,1) (f,0)} (f);
\path[tr] (c9) edge node[below right, text width=1cm] 
  { (d,0) (e,0) (f,1)} (f);

\paths{0}{1}
\paths{1}{2}
\paths{2}{3}
\paths{3}{4}
\paths{4}{5}

\paths{6}{7}
\paths{7}{8}
\paths{8}{9}

\path[tr,dashed] (a5) edge (a6);
\path[tr,dashed] (b5) edge (b6);
\path[tr,dashed] (c5) edge (c6);

\end{tikzpicture}

\caption{
    \ioautomaton{} used in the proof of Theorem~\ref{th:layered}.
}
\label{fig:base3}
\end{figure} 

\layered*

\begin{proof}
  Consider an instance $\varphi$ of \oneinthree{}, with a set of variables 
  $V = \set{x_0,\dots,x_{m-1}}$ with $m \geq 1$, 
  and a set of clauses $C \subseteq V^3$.
  Without loss of generality, we assume that $(x_i,x_j,x_k) \in C$
  implies $i < j < k$ (the variables that appear in a clause are ordered).
  
  We reduce \oneinthree{} to Problem~\ref{pb:layered} as follows.
  We define $\Sigma = \set{a,b,c}$ and $\Gamma = \set{0,1}$ (fixed alphabets),
  and $n = 9m+5$.
  
  Then, we define $E$ as the set of examples containing:
  \begin{itemize}
  \item $d a^{3m} d \zip 0^{3m+1}1$,
  \item $d a^{3m} e \zip 0^{3m+2}$,
  \item $d a^{3m} f \zip 0^{3m+2}$, \vspace{1ex}
  \item $e a^{3m} d \zip 0^{3m+2}$,
  \item $e a^{3m} e \zip 0^{3m+1}1$,
  \item $e a^{3m} f \zip 0^{3m+2}$, \vspace{1ex}
  \item $f a^{3m} d \zip 0^{3m+2}$,
  \item $f a^{3m} e \zip 0^{3m+2}$,
  \item $f a^{3m} f \zip 0^{3m+1}1$, \vspace{1ex}
  \item $d a^i c a^{3m-1-i} f \zip 0^{3m+2}$ for $0 \leq i < 3m$,
  \item $e a^i c a^{3m-1-i} d \zip 0^{3m+2}$ for $0 \leq i < 3m$,
  \item $f a^i c a^{3m-1-i} e \zip 0^{3m+2}$ for $0 \leq i < 3m$, \vspace{1ex}
  \item $d a^{3i} ccc a^{3(m-1-i)} d \zip 0^{3m+1}1$, for $0 \leq i < m$,
  \item $e a^{3i} ccc a^{3(m-1-i)} e \zip 0^{3m+1}1$, for $0 \leq i < m$,
  \item $f a^{3i} ccc a^{3(m-1-i)} f \zip 0^{3m+1}1$, for $0 \leq i < m$,
    \vspace{1ex}
  \item $d a^{3i} (caa) a^{3(j-i-1)} (caa) a^{3(k-j-1)} (caa) a^{3(m-k-1)} e
    \zip 0^{3m+1}1$ for $(x_i,x_j,x_k) \in C$.
  \end{itemize}
  
  We prove that $\varphi$ has a satisfying assignment if and only if 
  there exists a $(3m+2)$-layered and $(3m+2)$-total \ioautomaton{} $A$, consistent 
  with $E$, and with (at most) $n$ states.
  
  $(\Rightarrow)$
  Let $f: V \rightarrow \set{\bot,\top}$ be a satisfying assignment for 
  $\varphi$.
  We define $A = (\Sigma\times\Gamma,Q,\initState,\delta,F)$ following 
  Figure~\ref{fig:base3} with 
  \[
    Q = \set{ p_i,q_i,r_i,p_i',q_i',r_i',p_i'',q_i'',r_i''\ |\ 0 \leq i < m  }
   \cup \set{\initState,p_m,q_m,r_m,q_f} 
  \] 
  and $F = \set{q_f}$. $A$ has $9m + 5$ states. 
  
  Then, for each $x_i \in V$ for $0 \leq i < m$, if $f(x_i) = \top$, we add 
  nine transitions in $\delta$, called \emph{forward transitions}:
  \begin{itemize}
  \item $(p_i, (c,0), q_i')$,
  \item $(q_i, (c,0), r_i')$,
  \item $(r_i, (c,0), p_i')$, \vspace{1ex}
  \item $(p_i', (c,0), q_i'')$,
  \item $(q_i', (c,0), r_i'')$,
  \item $(r_i', (c,0), p_i'')$, \vspace{1ex}
  \item $(p_i'', (c,0), q_{i+1})$,
  \item $(q_i'', (c,0), r_{i+1})$,
  \item $(r_i'', (c,0), p_{i+1})$.
  \end{itemize}
  If $f(x_i) = \bot$, we add nice transitions as well, called 
  \emph{downward transitions}:
  \begin{itemize}
  \item $(p_i, (c,0), p_i')$,
  \item $(q_i, (c,0), q_i')$,
  \item $(r_i, (c,0), r_i')$, \vspace{1ex}
  \item $(p_i', (c,0), p_i'')$,
  \item $(q_i', (c,0), q_i'')$,
  \item $(r_i', (c,0), r_i'')$, \vspace{1ex}
  \item $(p_i'', (c,0), p_{i+1})$,
  \item $(q_i'', (c,0), q_{i+1})$,
  \item $(r_i'', (c,0), pr_{i+1})$.
  \end{itemize}
  Note that the definitions of forward and downward transitions are similar, 
  but different from the notions of forward and looping transitions introduced in
  the proofs of Theorems~\ref{th:fixedstates-oneexample} 
  and~\ref{th:fixedalphabet-oneexample}.
  
  $A$ is $(3m+2)$-layered \ioautomaton{} as it only accepts words of length $3m
  + 2$. Moreover, we can verify that $A$ is consistent with all the examples
  given in $E$. Finally, $A$ can be made $(3m+2)$-total by adding transitions
  which do not affect the examples in $E$.
  
  $(\Leftarrow)$
  Let $A = (\Sigma\times\Gamma,Q,\initState,\delta,F)$ be a 
  $(3m+2)$-total and $(3m+2)$-layered \ioautomaton{},
  consistent with $E$, and with $n = 9m+5$ states.
  Without loss of generality, we let $Q = \set{q_0,q_1,\dots,q_{n-1}}$ with 
  $\initState = q_0$.
  Moreover, in layered automata, a (reachable) accepting state only accepts 
  one word: $\emptyseq$. Therefore, without loss of generality, we can assume 
  that $A$ has a unique final state, and $F = \set{q_f}$ for some
  $q_f \in Q$, and $q_f \neq \initState$.
  
  The approach is the same as in Theorems~\ref{th:fixedstates-oneexample} and 
  \ref{th:fixedalphabet-oneexample}.
  First, using the examples:
  \begin{itemize}
  \item $d a^{3m} d \zip 0^{3m+1}1$,
  \item $d a^{3m} e \zip 0^{3m+2}$,
  \item $d a^{3m} f \zip 0^{3m+2}$, \vspace{1ex}
  \item $e a^{3m} d \zip 0^{3m+2}$,
  \item $e a^{3m} e \zip 0^{3m+1}1$,
  \item $e a^{3m} f \zip 0^{3m+2}$, \vspace{1ex}
  \item $f a^{3m} d \zip 0^{3m+2}$,
  \item $f a^{3m} e \zip 0^{3m+2}$,
  \item $f a^{3m} f \zip 0^{3m+1}1$,
  \end{itemize}
  We deduce that $A$ must contains the states and transitions 
  given in Figure~\ref{fig:base3}.
  
  Second, we prove that each variable $x \in V$ must have either 
  forward transitions, or downward transitions. We prove this using 
  the examples:
  \begin{itemize}
  \item $d a^i c a^{3m-1-i} f \zip 0^{3m+2}$ for $0 \leq i < 3m$,
  \item $e a^i c a^{3m-1-i} d \zip 0^{3m+2}$ for $0 \leq i < 3m$,
  \item $f a^i c a^{3m-1-i} e \zip 0^{3m+2}$ for $0 \leq i < 3m$, \vspace{1ex}
  \item $d a^{3i} ccc a^{3(m-1-i)} d \zip 0^{3m+1}1$, for $0 \leq i < m$,
  \item $e a^{3i} ccc a^{3(m-1-i)} e \zip 0^{3m+1}1$, for $0 \leq i < m$,
  \item $f a^{3i} ccc a^{3(m-1-i)} f \zip 0^{3m+1}1$, for $0 \leq i < m$.
  \end{itemize}
  We then use this fact to define an assignment $f$ that assigns 
  $\top$ to variables that have forward transitions, and $\bot$ to variables 
  that have downward transitions.
  
  Finally, for each clause $(x_i,x_j,x_k) \in C$, and using 
  \[
    d a^{3i} (caa) a^{3(j-i-1)} (caa) a^{3(k-j-1)} (caa) a^{3(m-k-1)} e
    \zip 0^{3m+1}1 \in A,
  \] 
  we deduce that exactly one variable out of 
  $x_i$, $x_j$ and $x_k$ must have forward transitions, and conclude that 
  $f$ is a satisfying assignment for $\varphi$.
  
  We now give more details for each step of the proof.
  We first prove that $A$ must contains the states and transitions 
  given in Figure~\ref{fig:base3}.
  Consider the fact that $d a^{3m} d \zip 0^{3m+1}1 \in A$.
  Then, the word $d a^{3m} d \zip 0^{3m+1}1$ must be accepted by a run without 
  cycle
  (a layered automaton cannot have any cycle, otherwise it would accept word 
  of different length).
  Without loss of generality, we call the states along this run (all distinct):
  $
    \initState,p_0,p_0',p_0'',\dots,p_{m-1},p_{m-1}',p_{m-1}'',p_m,q_f
  $.
  The examples $d a^{3m} e \zip 0^{3m+2}$ and $d a^{3m} f \zip 0^{3m+2}$
  ensure that there are transitions $(p_m,(e,0),q_f)$ as well 
  as $(p_m,(f,0),q_f)$.
  
  Then, consider the fact that $e a^{3m} d \zip 0^{3m+2} \in A$.
  Without loss of generality, we call the states along this run (all distinct):
  \[
    \initState,q_0,q_0',q_0'',\dots,q_{m-1},q_{m-1}',q_{m-1}'',q_m,q_f.
\]
  Next, we prove that the states 
  $q_0,q_0',q_0'',\dots,q_{m-1},q_{m-1}',q_{m-1}'',q_m$
  are all different from the states
  $p_0,p_0',p_0'',\dots,p_{m-1},p_{m-1}',p_{m-1}'',p_m$.
  Assume by contradiction that two of these states are equal,
  for instance 
  $p_0'' = q_0''$ (note that we cannot have two states from different levels 
  being equal, such as $q_0 = p_1$, as this would contradict our assumption 
  that $A$ is $l$-layered).
  Then all subsequent states must be equal as well, with $p_1 = q_1$,
  $p_1' = q_1'$, $p_1'' = q_1''$, \dots, and $p_m = q_m$.
  However, since there is a transition $(q_m,(d,0),q_f)$, this would implies
  that $d a^{3m} d \zip 0^{3m+2} \in A$, contradicting the fact that $A$ 
  is an \ioautomaton{}, as we know $d a^{3m} d \zip 0^{3m+1}1 \in A$,
  Similarly, by using the fact that $f a^{3m} d \zip 0^{3m+2} \in A$ and 
  introducing the states 
  $r_0,r_0',r_0'',\dots,r_{m-1},r_{m-1}',r_{m-1}'',r_m$
  along that run, we obtain the $A$ must be of the form 
  described in Figure~\ref{fig:base3}. Note that $A$ cannot have more 
  states that the one given in this figure, as we know that $A$ has at 
  most $9m+5$

  Then, we want to prove that for each $x_i \in V$, $0 \leq i < m$, 
  $x_i$ either has forward transitions or downward transitions, as defined
  in the first part of the proof.
  Let $i \in \set{0,\dots,m-1}$.
  Consider the facts that 
  $d a^{3i} c a^{3m-1-3i} f \zip 0^{3m+2} \in A$ and
  $d a^{3i+1} c a^{3m-2-3i} f \zip 0^{3m+2} \in A$ and
  $d a^{3i+2} c a^{3m-3-3i} f \zip 0^{3m+2} \in A$.
  After reading $d a^{3i} \zip 0^{3i+1}$,
  $A$ is in state $p_i$.
  Moreover, the only states from which 
  $a^{3m-1-3i} f \zip 0^{3m-3i}$ is accepted are 
  $p_i'$ and $q_i'$.
  Therefore, there must either exist 
  \begin{itemize}
  \item $(p_i,(c,0),p_i') \in \delta$, or
  \item $(p_i,(c,0),q_i') \in \delta$.
  \end{itemize}
  Similarly, because of $d a^{3i+1} c a^{3m-2-3i} f \zip 0^{3m+2} \in A$
  there must either exist
  \begin{itemize}
  \item $(p_i',(c,0),p_i'') \in \delta$, or
  \item $(p_i',(c,0),q_i'') \in \delta$.
  \end{itemize}
  And because of $d a^{3i+2} c a^{3m-3-3i} f \zip 0^{3m+2} \in A$,
  there must either exist a transition 
  \begin{itemize}
  \item $(p_i'',(c,0),p_{i+1}) \in \delta$, or
  \item $(p_i'',(c,0),q_{i+1}) \in \delta$.
  \end{itemize}
  
  Using the examples
  $e a^{3i} c a^{3m-1-3i} d \zip 0^{3m+2} \in A$ and
  $e a^{3i+1} c a^{3m-2-3i} d \zip 0^{3m+2} \in A$ and
  $e a^{3i+2} c a^{3m-3-3i} d \zip 0^{3m+2} \in A$, and 
  $f a^{3i} c a^{3m-1-3i} e \zip 0^{3m+2} \in A$ and
  $f a^{3i+1} c a^{3m-2-3i} e \zip 0^{3m+2} \in A$ and
  $f a^{3i+2} c a^{3m-3-3i} e \zip 0^{3m+2} \in A$, we also deduce the 
  following.
  There should exist transitions
  \begin{itemize}
  \item $(q_i,(c,0),q_i') \in \delta$, or
  \item $(q_i,(c,0),r_i') \in \delta$,
  \end{itemize}
  and
  \begin{itemize}
  \item $(q_i',(c,0),q_i'') \in \delta$, or
  \item $(q_i',(c,0),r_i'') \in \delta$,
  \end{itemize}
  and
  \begin{itemize}
  \item $(q_i'',(c,0),q_{i+1}) \in \delta$, or
  \item $(q_i'',(c,0),r_{i+1}) \in \delta$,
  \end{itemize}
  and
  \begin{itemize}
  \item $(r_i,(c,0),r_i') \in \delta$, or
  \item $(r_i,(c,0),q_i') \in \delta$,
  \end{itemize}
  and
  \begin{itemize}
  \item $(r_i',(c,0),r_i'') \in \delta$, or
  \item $(r_i',(c,0),q_i'') \in \delta$,
  \end{itemize}
  and
  \begin{itemize}
  \item $(r_i'',(c,0),r_{i+1}) \in \delta$, or
  \item $(r_i'',(c,0),q_{i+1}) \in \delta$.
  \end{itemize}
  
  Overall, there are $9$ choices, each with $2$ possibilities.
  Out of the $512$ combinations, we can verify that the examples
  \begin{itemize}
  \item $d a^{3i} ccc a^{3(m-1-i)} d \zip 0^{3m+1}1$,
  \item $e a^{3i} ccc a^{3(m-1-i)} e \zip 0^{3m+1}1$, and
  \item $f a^{3i} ccc a^{3(m-1-i)} f \zip 0^{3m+1}1$,
  \end{itemize}
  only allow $2$ outcomes.
  Either there are nine forward transitions:
  \begin{itemize}
  \item $(p_i, (c,0), q_i')$,
  \item $(q_i, (c,0), r_i')$,
  \item $(r_i, (c,0), p_i')$, \vspace{1ex}
  \item $(p_i', (c,0), q_i'')$,
  \item $(q_i', (c,0), r_i'')$,
  \item $(r_i', (c,0), p_i'')$, \vspace{1ex}
  \item $(p_i'', (c,0), q_{i+1})$,
  \item $(q_i'', (c,0), r_{i+1})$,
  \item $(r_i'', (c,0), p_{i+1})$,
  \end{itemize}
  or there are nine downward transitions:
  \begin{itemize}
  \item $(p_i, (c,0), p_i')$,
  \item $(q_i, (c,0), q_i')$,
  \item $(r_i, (c,0), r_i')$, \vspace{1ex}
  \item $(p_i', (c,0), p_i'')$,
  \item $(q_i', (c,0), q_i'')$,
  \item $(r_i', (c,0), r_i'')$, \vspace{1ex}
  \item $(p_i'', (c,0), p_{i+1})$,
  \item $(q_i'', (c,0), q_{i+1})$,
  \item $(r_i'', (c,0), pr_{i+1})$.
  \end{itemize}
  We define the assignment $f$ that assigns 
  $\top$ to variables that have forward transitions, and 
  $\bot$ to variables that have downward transitions.
  Let $(x_i,x_j,x_k) \in C$. 
  We know $d a^{3i} (caa) a^{3(j-i-1)} (caa) a^{3(k-j-1)} (caa) a^{3(m-k-1)} e
    \zip 0^{3m+1}1$.
    The only state from which the word 
    $a^{3(m-k-1)} e \zip 0^{3(m-k-1)+1}$ is accepted is $q_{k+1}$.
    Therefore, exactly one of the variables $x_i$, $x_j$ and $x_k$ must have
  forward transitions, while the two others must have downward transitions.
  This concludes the proof that $f$ is a satisfying assignment for $\varphi$.
  \end{proof}

\section{Proof of Lemma~\ref{lemma:sampleio}}
\label{app:sampleio}

\sampleio*

\begin{proof} 
  
  We define $T = (\Sigma \times \Gamma,Q^T,\initState^T,\delta^T,F^T)$
  to be a tree-shaped (partial) \ioautomaton{} consistent with $E$, as follows:
  \begin{itemize}
  \item $Q^T$ is the set of all prefixes of $E$,
  \item $\initState^T = \emptyseq$,
  \item $\delta^T = 
    \set{(q_1,(a,b),q_2)\ |\ q_1,q_2 \in E \land q_2 = q_1 \cdot (a,b) }$,
  \item $F^T = E$.
  \end{itemize}
  
  By construction, $T$ has at most $1 + \sum_{w \in E} |w|$ states.
  
  Let $P = \prefixes{\dom(E)} \subseteq \Sigma^*$ 
    be the set of all prefixes of $\dom(E)$.
  For each $u \in P$, we choose $v \in \Gamma^*$ as follows:
  \begin{itemize}
  \item if $u \in \dom(E)$, choose $v$ as the unique word such that 
    $u \zip v \in E$,
  \item 
    otherwise, choose any $v$ such that $u \zip v \in \prefixes{E}$.
  \end{itemize}
  We denote by $P' \subseteq \prefixes{E}$ the set of pairs $(u,v)$ 
  where $u \in P$ and $v$ is the corresponding word, chosen in the previous step.
  Let $b_0 \in \Gamma$ be a letter of the output alphabet.
  We define the automaton 
  $A = (\Sigma \times \Gamma,Q,\initState,\delta,F)$, 
  which is a total \ioautomaton{} consistent with $E$, as follows:
  \begin{itemize}
  \item $Q = Q^T \cup \set{q_f}$ where $q_f$ is a new state,
  \item $\initState = \initState^T$,
  \item 
    $\delta = \delta^T\ \cup$ \\
      \hspace*{2em}$\set{(q_f,(a,b_0),q_f)\ |\ a \in \Sigma}\ \cup$ \\
      \hspace*{2em}$\set{(q,(a,b_0),q_f)\ |\ 
        q \in P' \land 
        a \in \Sigma \land 
        \getin(q)\cdot a \notin P
       }$
  \item $F = P' \cup \set{q_f}$.
  \end{itemize}
  
  It remains to prove three things:
  (1) $A$ is an \ioautomaton{},
  (2) $A$ is total, and
  (3) $E \subseteq \getlang{A}$.
  
  \begin{enumerate}
  \item
      By construction, $A$ is a DFA. Let 
      $u \zip v_1 \in A$, and $u \zip v_2 \in A$,
      with $u \in \Sigma^*$ and $v_1,v_2 \in \Gamma^*$.
      Our goal is to prove that $v_1 = v_2$.
      We consider several cases:
      
      (a) $u \zip v_1$ and $u \zip v_2$ are both accepted in $q_f$:
      By construction of $A$, $q_f$ is a state from which a run can never 
      get out (a sink state). Consider the accepting run 
      of $u \zip v_1$ in $A$ and let $q_1 \in Q^T$ be the last state of $Q^T$
      before reaching $q_f$.
      There is a prefix $u_1 \zip v_1'$ of $u \zip v_1$ that corresponds
      to $q_1$.
      Similarly, let $q_2 \in Q^T$ be the last state of $Q^T$ in the 
      run of $u \zip v_2$ in $A$, and let $u_2 \zip v_2'$ be the prefix 
      of $u \zip v_2$ that corresponds to state $q_2$.
      Without loss of generality, we can assume that 
      $u_1$ is a prefix of $u_2$.
      
      Moreover, we prove that $u_1$ is in fact equal to $u_2$.
      Assume by contradiction that $u_1$ is a strict prefix of $u_2$,
      and let $u_2 = u_1 \cdot a \cdot u_1'$.
      Therefore, there is a transition from $q_1$ to $q_f$ whose input
      letter is $a$, which is not possible since $u_1 \cdot a \in P$.
      Therefore, $u_1 = u_2$.
      
      So far, we know $u_1 \zip v_1'$ goes to state $q_1$, and 
      $u_1 \zip v_2'$ goes to state $q_2$.
      By construction, the only transitions leading to $q_f$ are from 
      states of $P'$. So we have $q_1,q_2 \in P'$.
      We know $P'$ is a function relation, and only associates to each 
      word in $\Sigma^*$ at most one word in $\Gamma^*$.
      We deduce that $v_1' = v_2'$, and that $q_1 = q_2$.
      
      Since the runs then join $q_f$, where the only possible output
      letter is $b_0$, we deduce that $v_1 = v_2$.
  
      (b) $u \zip v_1$ is accepted in $q_f$, while $u \zip v_2$ is accepted 
          in $P'$
      (the case where $v_1$ and $v_2$ are interchanged is symmetrical):
      Consider the accepting run 
      of $u \zip v_1$ in $A$ and let $q_1 \in Q^T$ be the last state of $Q^T$
      before reaching $q_f$.
      Let  $u_1 \zip v_1'$ be the prefix of $u \zip v_1$ that 
      corresponds to $q_1$. 
      Let $u = u_1 \cdot a \cdot u_1'$ with $a \in \Sigma$
      and $u_1' \in \Sigma^*$.
      By construction of $q_1$, there is a transition 
      from $q_1$ to $q_f$ whose input letter is $a$. However, this is a
      contradiction, as $u_1 \cdot a \in P$.
      
      (c) $u \zip v_1$ and $u \zip v_2$ are both accepted in $P'$. $P'$ has been
          built as a functional relation, therefore we must have $v_1 = v_2$.
  
  \item Let $u \in \Sigma^*$. We want to prove that there exists $v \in
      \Gamma^*$ such that $u \zip v \in A$. Let $u = u' \cdot u''$ where $u'$ is
      the longest prefix of $u$ that belongs to $P$. Let $v' \in \Gamma^*$ be
      the unique word such that $u' \zip v' \in P'$. By defining $v = v' \cdot
      (b_0)^{|u''|}$, and by construction of $A$, we have $u \zip v \in A$.
  
  \item Since $A$ is obtained from $T$ by adding one state, some transitions,
      and by making some states accepting, we have $\getlang{T} \subseteq
      \getlang{A}$. Moreover, by construction of $T$, we have $E = \getlang{T}$,
      so we have $E \subseteq \getlang{A}$.
  \end{enumerate}
  \end{proof}
\section{Experiments Tables}
\label{app:tables}


\begin{table}
  \setlength\tabcolsep{0.2em}
  \begin{tabular}{c|c|c|c|c|c|c|c|c|c|c|c|c|c|c|c}
\backslashbox{i}{j} & 1 & 2 & 3 & 4 & 5 & 6 & 7 & 8 & 9 & 10 & 11 & 12 & 13 & 14 & 15 \\\hline
  1 & \pretty{0} & \pretty{0} & \pretty{0} & \pretty{0} & \pretty{0} & \pretty{0} & \pretty{0} & \pretty{0} & \pretty{0} & \pretty{0} & \pretty{1} & \pretty{1} & \pretty{1} & \pretty{1} & \pretty{0} \\\hline
  2 & \pretty{0} & \pretty{0} & \pretty{0} & \pretty{0} & \pretty{0} & \pretty{0} & \pretty{5} & \pretty{3} & \pretty{8} & \pretty{16} & \pretty{21} & \pretty{16} & \pretty{15} & \pretty{18} & \pretty{22} \\\hline
  3 & \pretty{0} & \pretty{0} & \pretty{0} & \pretty{3} & \pretty{1} & \pretty{10} & \pretty{16} & \pretty{32} & \pretty{24} & \pretty{36} & \pretty{35} & \pretty{33} & \pretty{44} & \pretty{44} & \pretty{41} \\\hline
  4 & \pretty{0} & \pretty{0} & \pretty{4} & \pretty{7} & \pretty{20} & \pretty{25} & \pretty{37} & \pretty{45} & \pretty{51} & \pretty{53} & \pretty{52} & \pretty{52} & \pretty{51} & \pretty{56} & \pretty{65} \\\hline
  5 & \pretty{0} & \pretty{0} & \pretty{6} & \pretty{20} & \pretty{35} & \pretty{46} & \pretty{57} & \pretty{63} & \pretty{59} & \pretty{64} & \pretty{67} & \pretty{62} & \pretty{60} & \pretty{60} & \pretty{64} \\\hline
  6 & \pretty{0} & \pretty{0} & \pretty{8} & \pretty{34} & \pretty{43} & \pretty{59} & \pretty{58} & \pretty{67} & \pretty{60} & \pretty{73} & \pretty{75} & \pretty{68} & \pretty{67} & \pretty{66} & \pretty{69} \\\hline
  7 & \pretty{0} & \pretty{0} & \pretty{17} & \pretty{37} & \pretty{61} & \pretty{65} & \pretty{70} & \pretty{70} & \pretty{81} & \pretty{76} & \pretty{78} & \pretty{72} & \pretty{75} & \pretty{73} & \pretty{75} \\\hline
  8 & \pretty{0} & \pretty{0} & \pretty{22} & \pretty{46} & \pretty{74} & \pretty{79} & \pretty{73} & \pretty{77} & \pretty{78} & \pretty{79} & \pretty{74} & \pretty{77} & \pretty{75} & \pretty{76} & \pretty{78} \\\hline
  9 & \pretty{0} & \pretty{0} & \pretty{22} & \pretty{63} & \pretty{67} & \pretty{76} & \pretty{86} & \pretty{80} & \pretty{78} & \pretty{79} & \pretty{82} & \pretty{83} & \pretty{84} & \pretty{82} & \pretty{80} \\\hline
  10 & \pretty{0} & \pretty{0} & \pretty{34} & \pretty{59} & \pretty{72} & \pretty{82} & \pretty{86} & \pretty{81} & \pretty{85} & \pretty{80} & \pretty{79} & \pretty{83} & \pretty{84} & \pretty{84} & \pretty{84} \\\hline
  11 & \pretty{0} & \pretty{0} & \pretty{36} & \pretty{73} & \pretty{82} & \pretty{86} & \pretty{83} & \pretty{85} & \pretty{85} & \pretty{89} & \pretty{88} & \pretty{86} & \pretty{91} & \pretty{82} & \pretty{83} \\\hline
  12 & \pretty{0} & \pretty{0} & \pretty{32} & \pretty{66} & \pretty{86} & \pretty{83} & \pretty{83} & \pretty{86} & \pretty{88} & \pretty{85} & \pretty{86} & \pretty{87} & \pretty{89} & \pretty{88} & \pretty{88} \\\hline
  13 & \pretty{0} & \pretty{0} & \pretty{41} & \pretty{83} & \pretty{85} & \pretty{85} & \pretty{89} & \pretty{87} & \pretty{89} & \pretty{85} & \pretty{93} & \pretty{89} & \pretty{88} & \pretty{89} & \pretty{89} \\\hline
  14 & \pretty{0} & \pretty{0} & \pretty{41} & \pretty{78} & \pretty{83} & \pretty{88} & \pretty{93} & \pretty{93} & \pretty{92} & \pretty{88} & \pretty{88} & \pretty{87} & \pretty{88} & \pretty{88} & \pretty{91} \\\hline
  15 & \pretty{0} & \pretty{0} & \pretty{51} & \pretty{83} & \pretty{87} & \pretty{87} & \pretty{88} & \pretty{84} & \pretty{91} & \pretty{87} & \pretty{91} & \pretty{91} & \pretty{90} & \pretty{87} & \pretty{88} \\
  \end{tabular}
  \\\vspace{2ex}
  
  \caption{ In a given cell, the number represents, out of 100 random automata,
    how many we were able to reobtain using our algorithm, with a random sample
    with $i$ input/output examples of length $j$.}
  \label{table:experiments}
 
\end{table}

\end{document}